\documentclass[prb, 11pt,letterpaper,superscriptaddress,floatfix,nofootinbib,notitlepage]{revtex4-1}

\usepackage{textcomp}
\usepackage{amsmath}
\usepackage{amsfonts}
\usepackage{amssymb}
\usepackage{graphicx}
\usepackage{siunitx}
\usepackage{setspace}
\setcitestyle{super}
\usepackage[colorlinks, citecolor={blue}, linkcolor={black}]{hyperref}
\usepackage{pdfpages}
\usepackage{pgffor}

\makeatletter
\AtBeginDocument{\let\LS@rot\@undefined}
\makeatother

\begin{document}
\raggedbottom

\author{Jeong Min Park}
\thanks{These authors contributed equally}
\affiliation{Department of Physics, Massachusetts Institute of Technology, Cambridge, Massachusetts 02139, USA}

\author{Yuan Cao}
\affiliation{Department of Physics, Massachusetts Institute of Technology, Cambridge, Massachusetts 02139, USA}

\author{$^{\!\!\!\!,\ \!*,\ \!\dagger}$ Kenji Watanabe}
\author{Takashi Taniguchi}
\affiliation{National Institute for Materials Science, Namiki 1-1, Tsukuba, Ibaraki 305-0044, Japan}

\author{Pablo Jarillo-Herrero}
\email{caoyuan@mit.edu}\email{pjarillo@mit.edu}
\affiliation{Department of Physics, Massachusetts Institute of Technology, Cambridge, Massachusetts 02139, USA}

\title{Flavour Hund's Coupling, Correlated Chern Gaps, and Diffusivity in Moir\'e Flat Bands}

\maketitle

\textbf{
Interaction-driven spontaneous symmetry breaking lies at the heart of many quantum phases of matter. In moir\'e systems, broken spin/valley `flavour' symmetry in flat bands underlies the parent state out of which ultimately correlated and topological ground states emerge\cite{cao_correlated_2018,cao_unconventional_2018,chen_evidence_2019, yankowitz_tuning_2019, lu_superconductors_2019, sharpe_emergent_2019, serlin_intrinsic_2020,  chen_tunable_2020,wong_cascade_2020,zondiner_cascade_2020}. However, the microscopic mechanism of such flavour symmetry breaking and its connection to the low-temperature many-body phases remain to be understood. Here, we investigate the symmetry-broken many-body ground state of magic angle twisted bilayer graphene (MATBG) and its nontrivial topology using simultaneous thermodynamic and transport measurements. We directly observe flavour symmetry breaking as a pinning of the chemical potential $\mu$ at all integer fillings of the moir\'e superlattice, highlighting the importance of flavour Hund's coupling in the many-body ground state. The topological nature of the underlying flat bands is manifested upon breaking time-reversal symmetry, where we measure energy gaps corresponding to Chern insulator states with Chern numbers $C=3,2,1$ at filling factors $\nu=1,2,3$, respectively, consistent with flavour symmetry breaking in the Hofstadter's butterfly spectrum of MATBG. Moreover, our concurrent measurements of resistivity and chemical potential allow us to obtain the temperature dependence of the charge diffusivity of MATBG in the strange metal regime\cite{cao_strange_2020}, a quantity previously explored only in ultracold atom systems\cite{brown_bad_2019}. Our results bring us one step closer to a unified framework for understanding interactions in the topological bands of MATBG, both in the presence and absence of a magnetic field.}

In condensed matter systems with flat electronic bands, the Coulomb interaction between electrons can easily surpass their kinetic energy and give rise to a variety of exotic quantum phases, ranging from Mott insulators to quantum spin liquids \cite{lee_doping_2006, balents_spin_2010}. In this strongly correlated regime, electrons may spontaneously order themselves to minimize the total Coulomb energy at the cost of increasing their kinetic energies, leading to the breaking of certain symmetries. Such symmetry-broken states can occur at a relatively high energy scale and act as a parent state for phases that appear at lower energy scales, such as superconductivity. Furthermore, when there is nontrivial topology in the system, the interplay between strong correlations and the underlying topology could lead to novel phases of matter, such as the fractional quantum Hall state\cite{stormer_fractional_1999}. Understanding the physics behind this interplay could guide us in designing next-generation strongly-correlated topological quantum materials.

Magic-angle twisted bilayer graphene (MATBG) serves as a unique platform to investigate interaction driven phenomena in a highly tunable flat-band system. When two layers of monolayer graphene (MLG) are stacked with a small twist angle of $\theta\sim\SI{1.1}{\degree}$, the interlayer hybridization in the resulting moir\'e superlattice renormalizes the Fermi velocity of the Dirac electrons and creates flat bands at low energies \cite{li_observation_2010, suarez_morell_flat_2010, bistritzer_moire_2011, lopes_dos_santos_continuum_2012}. In this regime, a plethora of exotic correlated and topological phenomena have been experimentally demonstrated, including correlated insulator states, superconductivity, and the quantum anomalous Hall effect\cite{cao_correlated_2018,cao_unconventional_2018, yankowitz_tuning_2019, lu_superconductors_2019, sharpe_emergent_2019,serlin_intrinsic_2020}. Scanning tunneling microscopy (STM) experiments have directly shown the significance of on-site Coulomb interactions in MATBG \cite{xie_spectroscopic_2019, kerelsky_maximized_2019,jiang_charge_2019, choi_electronic_2019}. More recently, scanning single-electron transistor and further STM measurements have suggested that the Coulomb interactions induce phase transitions that break the spin/valley symmetry \cite{wong_cascade_2020, zondiner_cascade_2020}. Despite significant experimental and theoretical progress, the microscopic picture that underlies the broken symmetry states in MATBG and their possible connections to the correlated phases and unusual superconductivity is still far from being complete and requires investigation.

\begin{figure}
\includegraphics[width=1\textwidth]{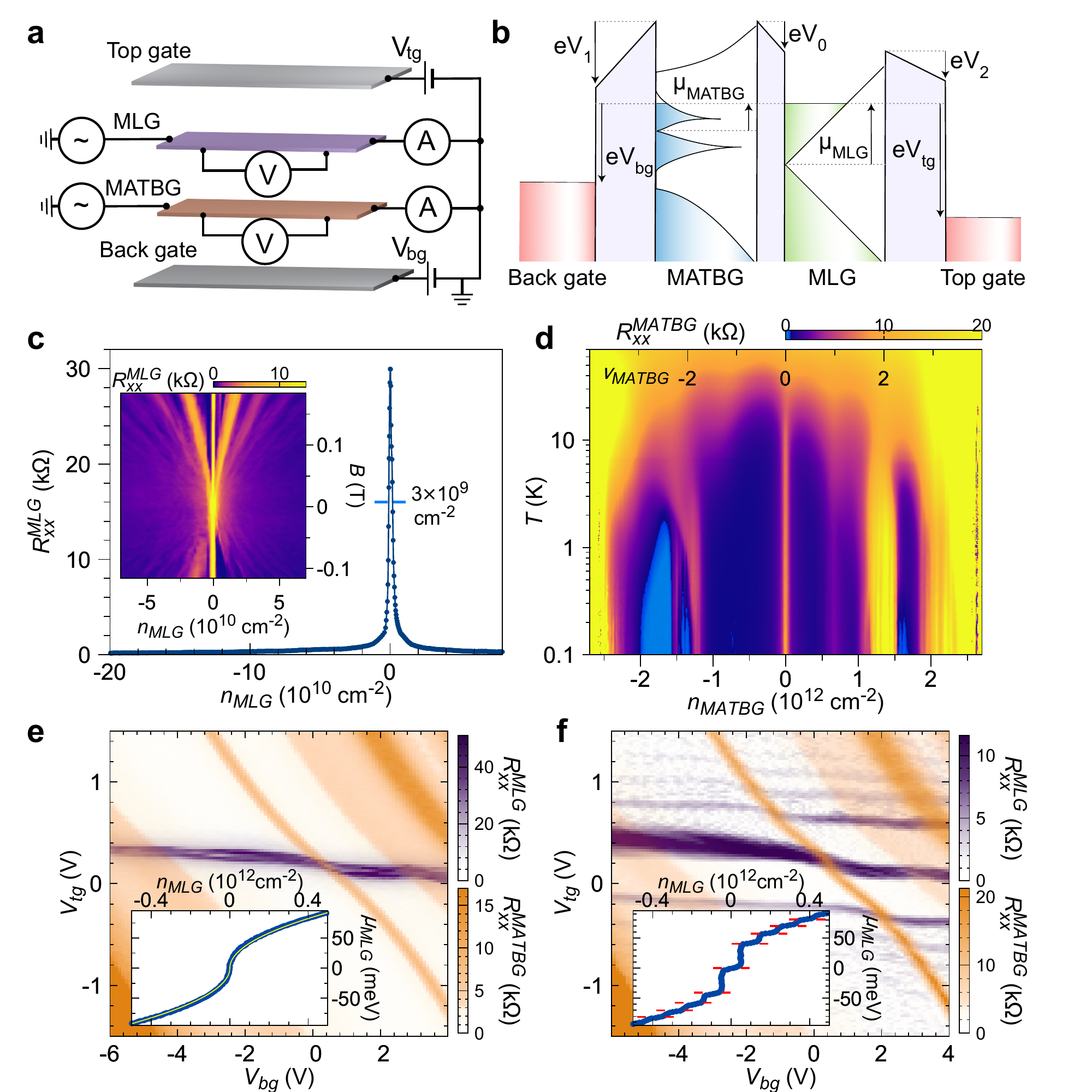}
\caption{\small Device structure and demonstration of chemical potential measurement. (a) Schematics of the measurement technique. MATBG and monolayer graphene (MLG) are separated by a thin ($\sim\SI{1}{nm}$) h-BN spacer and dual-gated. We simultaneously measure the resistance of MATBG and MLG. (b) Band diagram of the heterostructure, showing the relationship between the chemical potentials of MATBG ($\mu_\mathrm{MATBG}$) and MLG ($\mu_\mathrm{MLG}$), the back gate voltage $V_{bg}$ and top gate voltage $V_{tg}$, and the electrostatic potential drops $V_0$, $V_1$, and $V_2$. $e$ is the electron charge. (c) Transport characterization of MLG, showing a sharp resistance $R_{xx}^\mathrm{MLG}$ peak versus MLG carrier density $n_\mathrm{MLG}$, with full width half maximum of less than \SI{3e9}{\per\centi\meter\squared}. Inset: Landau fan diagram ($R_{xx}^\mathrm{MLG}$ versus $n_\mathrm{MLG}$ and magnetic field $B$) in MLG, which shows that the Landau levels become visible already at \SI{+-0.03}{\tesla}. (d) Transport characterization of MATBG. The twist angle of MATBG is $\theta=\SI{1.07+-0.03}{\degree}$. We find correlated states at filling factors $\nu_\mathrm{MATBG}=1, \pm2, 3$, as well as superconducting domes (blue) at $-2-\delta$ and $+2+\delta$, respectively. (e-f) Combined plot of the resistance of MLG and MATBG, represented by purple and orange colour scales, respectively, and overlaid in the same axes. As a proof of principle, we use the charge neutrality point (CNP) of MATBG (orange diagonal feature) to probe the chemical potential of MLG, at (e) $B=0$ and (f) $B_\perp=\SI{1}{\tesla}$. The horizontal purple stripes are the resistive features in MLG. From the CNP of MATBG, we extract the chemical potential $\mu_\mathrm{MLG}$ versus density $n_\mathrm{MLG}$, which is shown in the insets of (e-f). The white line in the inset of (e) is a fit to $\mu_\mathrm{MLG}=\hbar v_F \sqrt{\pi |n_\mathrm{MLG}|} \mathrm{sgn}(n_\mathrm{MLG})$. The red ticks in the inset of (f) denote the expected Landau level energies $\pm v_F\sqrt{2e\hbar B|N|}$, where $v_F=\SI{1.12e6}{\meter\per\second}$ and $N$ is an integer.}
\end{figure}

Here we study the interplay between interaction-driven symmetry breaking and nontrivial topology in the flat bands of MATBG by directly measuring the combined thermodynamic and transport properties of its many-body ground state. For this, we use a unique technique\cite{kim_direct_2012, lee_chemical_2014} that involves a MLG probe layer in close proximity to MATBG to sense the chemical potential of MATBG at different charge densities, temperatures, and magnetic fields. Measuring chemical potential or compressibility is a class of techniques\cite{feldman_unconventional_2012, tomarken_electronic_2019} complementary to the spectroscopic techniques that probe the excitation spectra, such as tunneling or photoemission spectroscopy. We find that the chemical potential distinctively reaches local extrema when the number of electrons per moir\'e unit cell ($\nu$) is close to integers $\pm1, \pm2, \pm3$ and $\pm4$. We show that these results can be naturally explained in the framework of spin/valley `flavour' symmetry breaking, but where in addition to Coulomb repulsion\cite{kerelsky_maximized_2019, zondiner_cascade_2020, wong_cascade_2020, choi_electronic_2019}, we need to consider the intra-flavour Hund's coupling. The latter interaction results in the pinning of the chemical potential, favouring single flavour occupancy in a way analogous to the Hund's rule for spin alignment in multi-electron atoms. Surprisingly, the response of the chemical potential to the in-plane magnetic field indicates that a finite in-plane magnetization develops at $\nu=\pm1,\pm2,\pm3$ at finite magnetic field, including the half-filling ($\nu=\pm2$), suggesting that the magnetic state of the $\nu=\pm2$ correlated states might be more intricate than previously thought \cite{cao_correlated_2018,yankowitz_tuning_2019,wu_chern_2020}. Furthermore, the nontrivial topology of the MATBG flat bands is revealed when the time-reversal symmetry is broken by applying a perpendicular magnetic field\cite{tomarken_electronic_2019, wu_chern_2020, nuckolls_strongly_2020, saito_hofstadter_2020, das_symmetry_2020}. We directly observe and measure the size of the correlated gaps with Chern numbers 3, 2, and 1, and demonstrate that all the experimentally observed states can be explained in the unified framework of symmetry broken Hofstdater spectrum \cite{hofstadter_energy_1976, macdonald_butterfly_2011}. With combined chemical potential and transport measurements, we compare the temperature dependence of resistivity, compressibility, and charge diffusivity in the high-temperature regime of MATBG where linear resistivity-temperature behaviour is observed \cite{cao_strange_2020, polshyn_large_2019}. Our data suggest that the diffusivity of MATBG is close to a diffusivity bound\cite{hartnoll_theory_2015} proposed in the incoherent transport regime.

Figure 1a illustrates our experimental scheme. The MATBG is separated from a monolayer graphene (MLG) layer by an ultrathin layer of h-BN ($\sim \SI{1}{\nano\meter}$). The MATBG layer is fabricated by a `laser-cut \& stack' method (see Supplementary Information), which reduces the strain compared to our previous `tear \& stack' method. We use the top gate voltage $V_{tg}$ and back gate voltage $V_{bg}$ to control the densities in MLG and MATBG, and measure the transport properties of the two layers simultaneously. In this setup, direct probing of the chemical potential $\mu$ of one layer is achieved by sensing the screening of electric field from the gates by the other layer \cite{kim_direct_2012, lee_chemical_2014}. From the band alignment diagram shown in Fig. 1b, we can deduce the relationship between $(V_{tg}, V_{bg})$ and $(n_\mathrm{MLG}, \mu_\mathrm{MLG}, n_\mathrm{MATBG}, \mu_\mathrm{MATBG})$, the latter being density ($n$) and chemical potential ($\mu$) in MLG and MATBG, respectively (see Supplementary Information for details). In particular, when one layer is at the charge neutrality point, \emph{e.g.} $n_\mathrm{MLG}=0$, the chemical potential of the other layer ($\mu_\mathrm{MATBG}$) is directly proportional to one of the gate voltages, which in this case is given by $\mu_\mathrm{MATBG}=-(eC_{tg}/C_{i})V_{tg}$, where $C_{tg}$ and $C_i$ are the geometric capacitances per unit area of the top and middle h-BN dielectrics, respectively.

The MLG layer used in our experiments has very low disorder $<\SI{3e9}{\per\centi\meter\squared}$ and excellent field-effect mobility $>\SI{300000}{\centi\meter\squared\per\volt\per\second}$ (Fig. 1c). The Landau levels start developing from magnetic fields as low as $\pm\SI{30}{\milli\tesla}$. The MATBG layer has a twist angle of $\theta=\SI{1.07+-0.03}{\degree}$, and exhibits correlated states at integer filling factors $\nu_\mathrm{MATBG}=4n_\mathrm{MATBG}/n_s=+1,\pm2,+3$ of the flat bands ($n_s=8\theta^2/\sqrt{3}a^2$ is the superlattice density of TBG and $a=\SI{0.246}{\nano\meter}$ is the lattice constant of graphene), as well as superconducting states at both $\nu=-2-\delta$ and $+2+\delta$, where $\delta$ is a small change in filling. The superconducting transition temperature $T_c$ reaches as high as \SI{2.7}{\kelvin} for $\nu=-2-\delta$, as determined from the \SI{50}{\percent} normal state resistance (see Extended Data Figure 1). The outstanding quality of both the MATBG and the MLG probe layers creates an exceptional platform to study the underlying physics in MATBG.

To demonstrate the measurement principle, we first measure the chemical potential and resistivity of MLG with and without a magnetic field. A key advantage of this experimental technique is that we can simultaneously measure electronic transport in both layers and accurately correlate the chemical potential to the transport features. Fig. 1e and f show the resistance of MATBG and MLG as a function of $V_{tg}$ and $V_{bg}$ at $B_{\perp}=\SI{0}{\tesla}$ and $B_{\perp}=\SI{1}{\tesla}$, respectively. $\mu_\mathrm{MLG}$ as a function of $n_\mathrm{MLG}$ is obtained by tracking the charge neutrality of MATBG (see inset of Fig. 1e and Supplementary Information for the conversion formulae). From these extracted values, we determine the Fermi velocity to be $v_F = \SI{1.12e6}{\meter\per\second}$ by fitting to $\mu_\mathrm{MLG}=\hbar v_F \sqrt{\pi |n_\mathrm{MLG}|} \mathrm{sgn}(n_\mathrm{MLG})$. In a magnetic field $B_\perp=\SI{1}{\tesla}$, the spectrum of MLG is quantized into discrete Landau levels. The extracted chemical potential in these Landau levels fits well to the Landau level spectrum of MLG, where the energy of the $N$-th Landau level is $\pm v_F\sqrt{2e\hbar B|N|}$. Our technique can thus determine the chemical potential of either layer with a sensitivity of $\lesssim$ \si{\milli\electronvolt}.

The chemical potential of MATBG is shown in Fig. 2a. Hereafter we will simply use $n$ ($\nu$) and $\mu$ to denote $n_\mathrm{MATBG}$ ($\nu_\mathrm{MATBG}$) and $\mu_\mathrm{MATBG}$. We show the data as a function of the MATBG filling factor $\nu$ and $V_{tg}$, which is directly proportional to the chemical potential of MATBG $\mu$ if one tracks the charge neutrality point of MLG (shown as the green curve). The resistance data of MLG (purple) and MATBG (orange) are overlaid for qualitative comparison of features, and the gray dash lines indicate the integer filling factors $\nu=0,\pm1,\pm2,\pm3$ of MATBG, which correspond to filling 0, 1, 2, or 3 electrons (holes) per moir\'e unit cell, respectively. The rate that $\mu$ increases with $n$ (or $\nu$), known as the inverse electronic compressibility $\chi^{-1}=d\mu/dn$, is inversely proportional to the density of states (DOS) for a non-interacting system. Around the charge neutrality ($\nu=0$), $\mu$ rises quickly with $\nu$, consistent with a minimal DOS at the Dirac point. However, once we start filling electrons into the flat band, its rate of increase decreases quickly and $\mu$ reaches a local maximum around $\nu=0.6$. Surprisingly, it then starts decreasing, exhibiting a negative $\chi^{-1}$, \cite{eisenstein_compressibility_1994} and it gets pinned at a local minimum around the integer filling $\nu=1$. Having this as a turning point, $\mu$ rises again until it hits the next maximum. This intriguing pinning behaviour repeats itself at each integer filling factor, including $\nu=4$ (see Fig. 2a inset). On the hole-doped side ($\nu<0$), the pinning behaviour of the chemical potential is opposite and weaker (i.e creates weak maxima in $\mu$). The total bandwidth estimated from the chemical potential at low temperatures is around $\sim$\SI{40}{\milli\electronvolt}, where both the electron and hole-doped sides contribute similarly. We also investigated the behaviour of $\mu$ while varying temperature from $\SI{2}{K}$ to $\SI{70}{K}$, as shown in Fig. 2d. The observed pinning behaviour persists prominently up to $\SI{20}{K}$. We note that the observation of a negative compressibility indicates that our system might be in a strong Coulomb frustration regime \cite{ortix_coulomb-frustrated_2008}, which acts to suppress macroscopic phase separation that may occur otherwise in an unconstrained system.

\begin{figure}[t]
\includegraphics[width=1\textwidth]{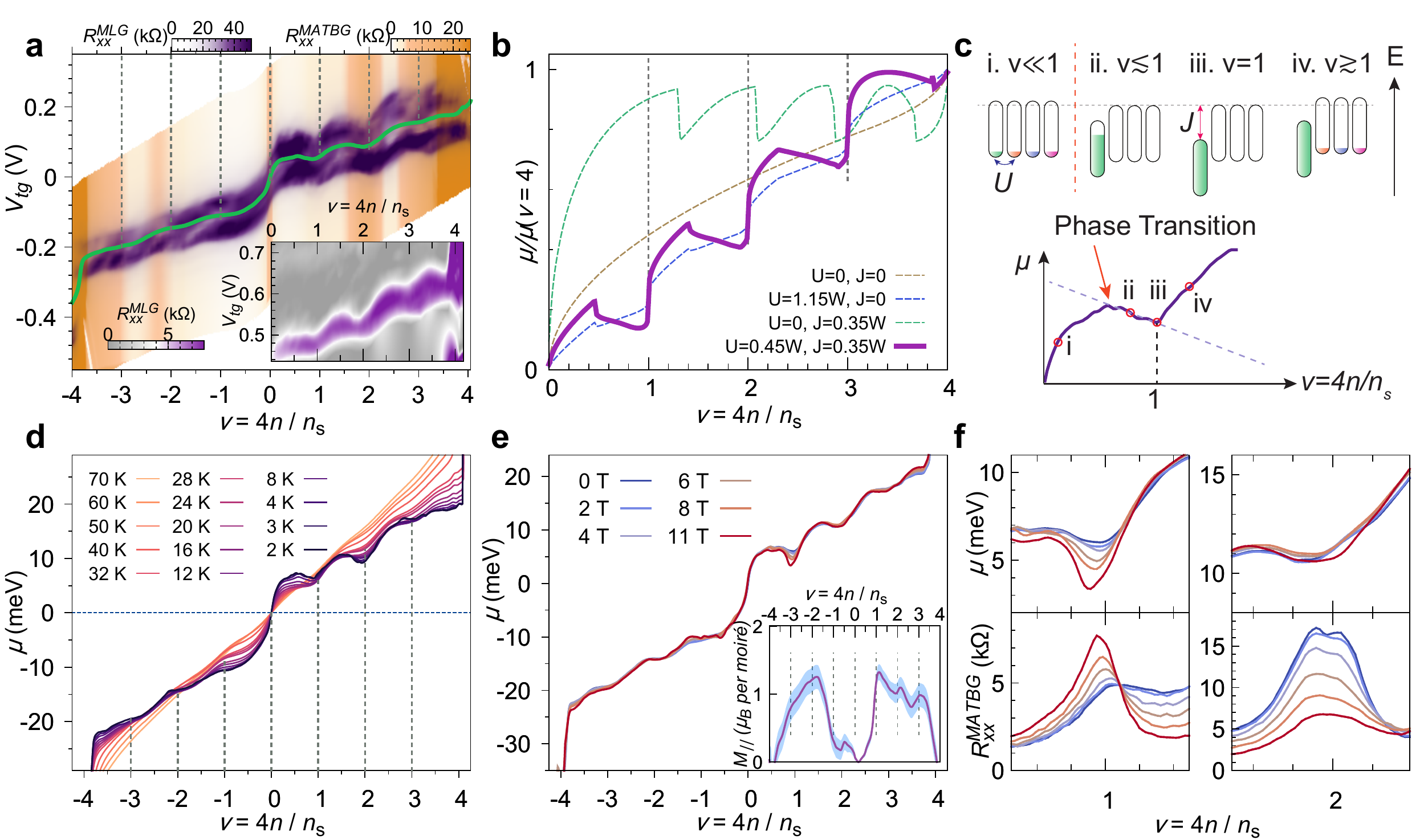}
\caption{\small Chemical potential of MATBG as a function of temperature and in-plane magnetic field. (a) Sensing the chemical potential of MATBG using the MLG charge neutrality point (CNP). Measurement taken at $B=\SI{0}{\tesla}$ and $T=\SI{4}{\kelvin}$. The green line shows the extracted chemical potential of MATBG. Gray dash lines mark the four filling factors of MATBG obtained from the Landau Fan diagram, which agree with the MATBG correlated states resistive features. The chemical potential is pinned at each filling factor, showing the stabilization of the state. The inset shows the same features probed by tracking the $N=1$ MLG LL at $B=\SI{0.7}{\tesla}$. (b) Mean-field estimate of the chemical potential with various Coulomb repulsion energy $U$ and exchange energy $J$ in units of the single-particle bandwidth $W\equiv1$. The experimental data are best explained qualitatively when both terms are nonzero. (c) Illustration of interaction-driven chemical potential stabilization at $\nu=1$. Chemical potential curve at $T=\SI{2}{\kelvin}$ near $\nu=1$ is shown. A phase transition associated with flavour symmetry breaking occurs before each integer filling factor (except $\nu=4$). The exchange energy $J$ stabilizes the filled flavour when the filling factor is close to one. (d) Temperature dependence of the chemical potential of MATBG from $T=\SI{2}{\kelvin}$ to $T=\SI{70}{\kelvin}$, probed with the MLG CNP. Clear pinning behaviour at integer filling factors persists up to $T=\SI{20}{\kelvin}$. (e) In-plane magnetic field, $B_{\parallel}$, dependence of the chemical potential of MATBG at $T=\SI{4}{\kelvin}$ and $B_{\perp}=\SI{0.7}{\tesla}$, probed with the $N=1$ MLG LL. The pinning in chemical potential around odd filling factors $\nu = \pm1$ gets intensified as $B_{\parallel}$ is applied, whereas those at filling factors $\pm2$ do not display significant change. Inset: Magnetization $M_{\parallel}$ in units of Bohr magneton $\mu_{B}$ per moir\'e unit cell, which shows that all states at $\nu=\pm1,\pm2,\pm3$ are magnetized in an in-plane field. Error bands (blue) correspond to \SI{95}{\percent} confidence interval. (f) Zoom-in of the chemical potential (top) and transport resistance (bottom) aligned for comparison, shown for $\nu=+1, +2$.}
\end{figure}

The pinning of $\mu$ at all integer $\nu$ is reminiscent of the stabilization of electronic shells in atoms when they are half-filled or full-filled, which is known as Hund's rule for maximum spin multiplicity. The physical origin of the Hund's rule stems from the Coulomb exchange interaction between the electrons. Here in MATBG, we also find that the pinning behaviour of the chemical potential is naturally explained when both the on-site inter-flavour Coulomb repulsion energy $U$ and inter-site intra-flavour exchange energy $J$ are considered. We focus on the $\nu>0$ side in the following description. Figure 2b shows the chemical potential calculated with a mean-field model for different values of $U$ and $J$. Our model can reproduce qualitatively the experimentally measured chemical potential  only when both $U$ and $J$ are nonzero and of similar magnitude (purple solid curve), beyond the currently established understanding\cite{wong_cascade_2020, zondiner_cascade_2020,choi_electronic_2019}. We wish to emphasize the importance of the exchange energy in stabilizing the chemical potential by illustrating a possible mechanism for the $\nu=1$ case (see Fig. 2c). Near charge neutrality, as the density is increased, all four flavours are filled at the same rate. As $\nu$ starts to approach one, the Coulomb repulsion between different flavours starts to surpass the kinetic energy penalty of filling up only one flavour. As $\nu$ reaches a certain value (still below 1), a flavour-symmetry-breaking phase transition occurs and all electrons are transferred into a single flavour to minimize the Coulomb repulsion \cite{wong_cascade_2020, zondiner_cascade_2020}. From this phase transition point all the way to $\nu=1$, i.e. while a single flavour is being filled, the $U$ term does not have any contribution to the free energy, while the $J$ term decreases the total free energy as $\sim -J\nu^2$ (see Supplementary Information). This term decreases the chemical potential and results in a negative inverse compressibility $\chi^{-1}\propto \mathcal{D}^{-1}-2J$ ($\mathcal{D}$ is the single-particle DOS per flavour) when $2J>\mathcal{D}^{-1}$. At $\nu=1$, maximal stabilization by the exchange term $J$ is reached, and thus the pinning of $\mu$. Further increase in $\nu$ populates the other three empty flavours and it increases the chemical potential before the next phase transition occurs. We also note that the pinning of chemical potential on the hole-doped side occurs at slightly more negative values of $\nu$ compared to exact integers, which may be attributed to smaller $U/W$ and/or $J/W$ ratios on that side, where $W$ is the single-particle bandwidth (see Supplementary Information).

\begin{figure}
\includegraphics[width=\textwidth]{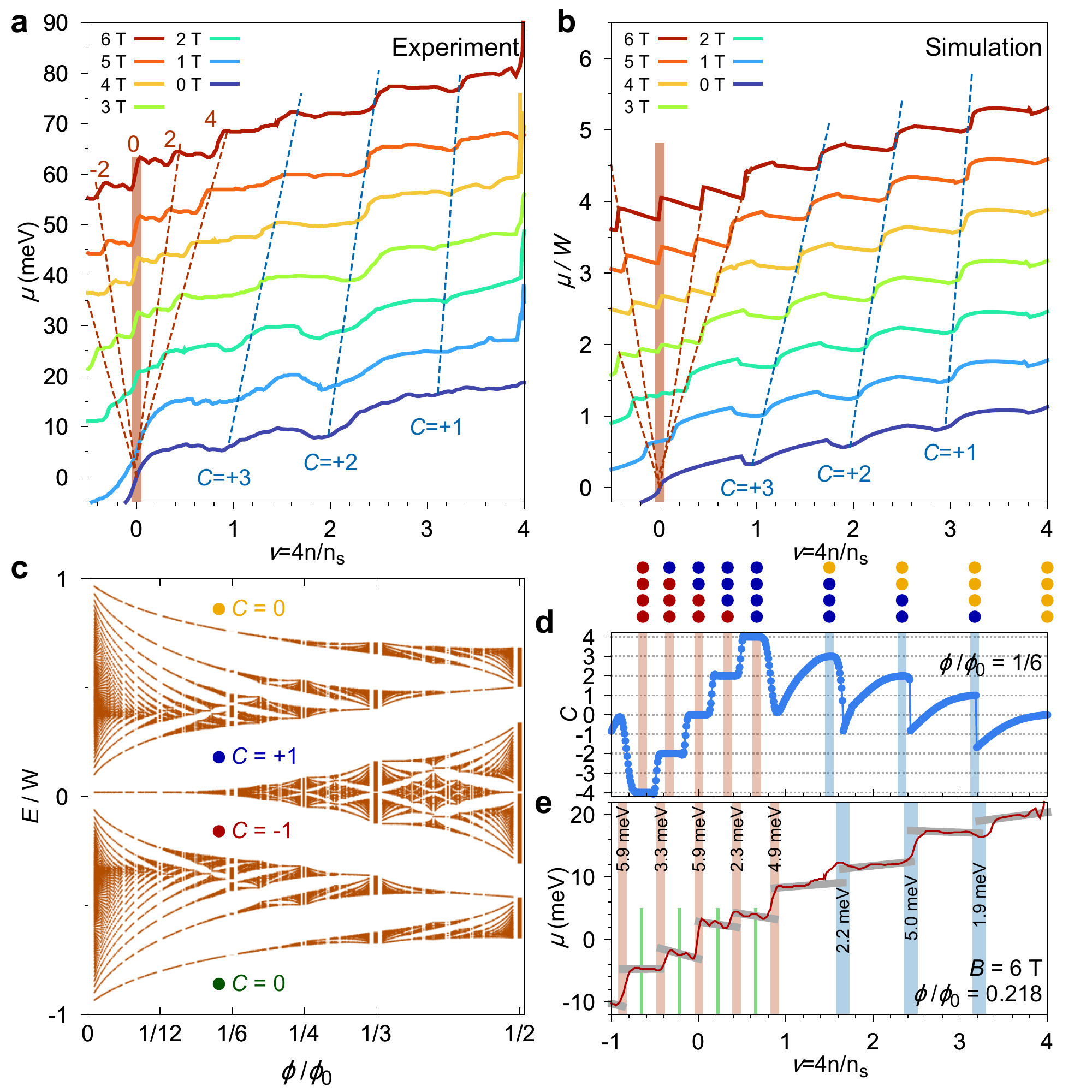}
\caption{\small Probing the correlated Chern gaps of MATBG in a perpendicular magnetic field. (a) Experiment and (b) simulation of the chemical potential versus $\nu$ in MATBG, at $B_\perp$ from zero to \SI{6}{\tesla}. $W$ is the bandwidth used in the simulation (see Supplementary Information). Near charge neutrality we find gaps that correspond to Landau level filling factors  $\nu_{LL}=0, \pm2,$ and $\pm4$, while the pinning of $\mu$ at $\nu=1,2,3$ shown in Fig. 2 evolves into topological gaps with Chern numbers $C=3,2,1$, respectively, as evident from their slope in magnetic field $dn/dB=C/\phi_0$, where $\phi_0$ is the flux quantum. (c) The Hofstadter's butterfly spectrum of TBG up to a flux per unit cell of $\phi_0/2$ (calculation shown for \SI{1.8}{\degree}, but spectrum is qualitatively similar for MATBG). The major gaps in the spectrum have Chern numbers of $C=0, -1, +1, 0$ per flavour, respectively. (d) Calculated total Chern number of TBG using the mean-field model with Coulomb repulsion and exchange interactions for a flux of $\phi_0/6$. The correct Chern number is reproduced, both in the Landau levels near the charge neutrality ($C=-4,-2,0,2,4$, indicated by red bars) and in the correlated Chern gaps ($C=3,2,1$, indicated by the blue bars). The dots above the plot show the configuration of the four flavours in each gap. The colouring scheme of the dots matches the ones shown in (c). Adding the Chern number of each flavour gives the total Chern number. (e) Extraction of energy gaps in the correlated spectrum of MATBG at $B_\perp=\SI{6}{\tesla}$. See Extended Data Figure 4 for their dependence on $B_\perp$.}
\end{figure}

To directly probe the magnetic properties of the correlated states, we measured the chemical potential in MATBG as a function of in-plane magnetic field up to \SI{11}{\tesla}. The results are shown in Fig. 2e and zoomed-in in Fig. 2f. In the lower panels of Fig. 2f, we also show the resistance of MATBG as measured by transport. At $\nu=\pm 1$, the pinning of the chemical potential is clearly strengthened by $B_{\parallel}$, as is the intensity of the transport resistance peak (see Methods and Extended Data Figure 2). These findings suggest that the states $\nu=\pm 1$ develop a spin-polarization in response to the magnetic field. To confirm this, we directly obtained the magnetization by integrating the Maxwell's relation $\left(\frac{\partial M_{\parallel}}{\partial \nu}\right)_{B_{\parallel},T} = -\left(\frac{\partial\mu}{\partial B_{\parallel}}\right)_{\nu,T}$, where $M_{\parallel}$ is the magnetization per moir\'e unit cell induced by the field\cite{zondiner_cascade_2020}. The inset of Fig. 2e shows $M_\parallel$ in units of Bohr magneton $\mu_B$ per moir\'e unit cell. We indeed find that the magnetization reaches a value on the order of one $\mu_B$ at $\nu=\pm1$, consistent with a spin-polarized state at finite field, which would indicate either a very soft paramagnetic state or a ferromagnetic state at zero field. We do not observe hysteresis in transport, which suggests it may be the former. The $\nu=\pm2$ states, on the other hand, have been speculated to be spin-unpolarized insulating states\cite{cao_correlated_2018,yankowitz_tuning_2019,wu_chern_2020}. However, we find that, while the resistance of these insulating states is indeed suppressed by the in-plane magnetic field (see Fig. 2f and Extended Data Figure 2), to our surprise the chemical potential measured at $\nu=\pm2$ does not show significant dependence on the in-plane magnetic field (Fig. 2f). Furthermore, $M_{\parallel}$ does not return to zero when $\nu$ is tuned from $\pm1$ to $\pm2$ (Fig. 2e inset). While the lack of dependence of $\mu$ on $B_\parallel$ at $\nu=\pm2$ can be partially captured by our theoretical model (see Supplementary Information), the persistence of magnetization near $\nu=\pm2$ is at odds with the finite-field spin-unpolarized state inferred from the suppression of the transport gap. These observations suggest that in an in-plane field the $\nu=\pm2$ gaps might select a ground state with nontrivial spin and/or valley texture, beyond simply occupying two flavours with opposite spins.  

Our experiment also puts constraints on the possible mechanism of superconductivity in MATBG. As shown in Extended Data Figure 5b, the superconducting dome lies in the region where $\chi^{-1}$ is high, with maximum $T_c$ corresponding to a maximum in $\chi^{-1}$. Since, in the non-interacting limit, $\chi$ is equal to the single-particle DOS,  a Bardeen-Cooper-Schrieffer (BCS) type superconductivity would be enhanced when the DOS is high and thus have the highest $T_{c}$ when $\chi^{-1}$ is low. Therefore, our observation of an opposite trend indicates that it is not easy to reconcile the superconductivity in MATBG with a weakly-coupled BCS theory. Future theories attempting to model the superconductivity in MATBG will likely need to take into account the importance of Coulomb interactions, including both repulsion and Hund's coupling, and the consequent phase transitions.

We now turn to the topological properties of MATBG. By measuring the chemical potential of MATBG in a perpendicular magnetic field, we can gain insight into the energy gaps in MATBG that result from the interplay between the topology of the Hofstadter spectrum and the Coulomb interactions, as suggested by recent experiments \cite{tomarken_electronic_2019, wu_chern_2020, nuckolls_strongly_2020, saito_hofstadter_2020,das_symmetry_2020}. The helical nature of the Dirac electrons in graphene endows each flat band of MATBG a Chern number of $C=\pm1$, which however is only explicitly manifested when the composite inversion-time reversal ($C_2\mathcal{T}$) symmetry is broken, either by alignment to the h-BN substrate (breaks $C_2$) or by applying a magnetic field (breaks $\mathcal{T}$). Fig. 3c shows the Hofstadter butterfly spectrum of TBG, where the topologically nontrivial gaps with $C=\pm1$ and the trivial gaps with $C=0$ are shown. The former gaps are smoothly connected to the Landau level gaps at  $\nu_{LL}=\nu/(\phi/\phi_0)=\pm4$ at low fields, where $\phi$ is the magnetic flux per unit cell and $\phi_0=h/e$ is the flux quantum. Without interactions, the only possible total Chern number in this picture is $C_{tot}=0, \pm4$, since all flavours are in the same gap. The Coulomb interactions among the flavours can cause their Chern numbers to be different, and can give rise to new hierachies of correlated Chern gaps. 

These topological gaps are directly observed in our chemical potential measurements, as shown in Fig. 3a. Near charge neutrality, we observe the quantum Hall gaps as steps in the chemical potential at the Landau level filling factors $\nu_{LL}=0, \pm2, \pm4$, whose positions evolve according to the Streda formula $dn/dB = \nu_{LL}/\phi_0$ \cite{streda_thermodynamic_1983}. The appearance of the Landau level gaps at $\nu_{LL}=0, \pm2$ indicates that the flavour symmetry is already broken. In the meantime, the extrema in the chemical potential at $\nu=1,2,3$ at $B_\perp=0$ evolve into topological gaps at $B_\perp=\SI{6}{\tesla}$. The topological nontriviality of these gaps is evident from the fact that their evolution follows the same Streda formula $dn/dB = C/\phi_0$ that indicates the total Chern number of $C=3,2,1$ associated with the states originally at $\nu=1,2,3$, respectively. 

The appearance of the broken-symmetry Landau levels and topological Chern gaps can be analyzed in a unified way using a correlated Hofstdater spectrum model\cite{hofstadter_energy_1976, macdonald_butterfly_2011}. We consider the single-particle DOS to be representative of the Hofstadter spectrum shown in Fig. 3c\cite{macdonald_butterfly_2011}, with possible Chern numbers $C=0, -1, +1, 0$ associated with the major gaps for each flavour, and add the mean-field Coulomb repulsion and exchange terms $U$ and $J$ in a similar manner as above. Using this model, we calculate the Chern number $C$ as a function of total filling $\nu$ and reproduce the experimentally observed sequence of $0, \pm2, \pm4$ at the charge neutrality Landau levels, and $3,2,$ and $1$ at densities $\nu=1+3\frac{\phi}{\phi_0}, 2+2\frac{\phi}{\phi_0},$ and $3+\frac{\phi}{\phi_0}$, respectively, as shown in Fig. 3d. In the top part, we also illustrate the contribution to the total Chern number from each flavour by colour-coded dots in accordance with Fig. 3c. By performing a similar calculation (see Supplementary Information), we can simulate the evolution of the chemical potential with the magnetic field, as shown in Fig. 3b. The remarkable similarity with the experimental data clearly indicates that this model captures the main features of the correlated spectrum of MATBG with and without a magnetic field.

A more quantitative analysis is performed on the chemical potential measured at $B_\perp=\SI{6}{\tesla}$, as shown in Fig. 3e. From the steps in $\mu$, we can directly extract the sizes of all energy gaps in the spectrum without relying on any temperature-dependent measurement. The Landau level gaps at $\nu_{LL}=-4,-2,0,2,$ and $4$ are $5.9, 3.3, 5.9, 2.3,$ and $\SI{4.9}{\milli\electronvolt}$, respectively. The small sizes of the gaps at $\nu_{LL}=\pm4$ translate to a vastly renormalized Fermi velocity of approximately $v_F\sim\SI{6e4}{\meter\per\second}$, consistent with other experimental probes in MATBG\cite{cao_correlated_2018, tomarken_electronic_2019}. The gaps at $\nu_{LL}=0, \pm2$, on the other hand, are broken-symmetry gaps created by the Coulomb interactions $U$ and $J$, and have similar sizes as those found in the $\nu_{LL}=0, \pm1$ broken-symmetry states in MLG\cite{young_spin_2012,zhao_symmetry_2010}. However, a fundamental difference with MLG is that the broken-symmetry gaps in MATBG have the same energy scale as the single-particle gaps at $\nu_{LL}=\pm4$, another manifestation of the fact that $U, J$ are on the same order as $W$. We also find clear evidence of smaller gaps at $\nu_{LL}=\pm1, \pm3$, which require further symmetry breaking than those discussed here (see Supplementary Information). Furthermore, the sizes of the topological Chern gaps at $\nu=1+3\frac{\phi}{\phi_0}, 2+2\frac{\phi}{\phi_0},$ and $3+\frac{\phi}{\phi_0}$ are extracted to be 2.2, 5.0 and \SI{1.9}{\milli\electronvolt} respectively. The larger gap at $\nu=2+2\frac{\phi}{\phi_0}$ is consistent with the fact that this state is more readily resolved in electronic transport experiments\cite{cao_correlated_2018,cao_unconventional_2018,yankowitz_tuning_2019,lu_superconductors_2019, wu_chern_2020, saito_hofstadter_2020}. Its difference with the gaps at $\nu=1+3\frac{\phi}{\phi_0}$ and $3+\frac{\phi}{\phi_0}$ might be attributed to the different magnetic ground state, with contributions from both orbital and spin degrees of freedom of the two fully filled flavours. These gaps have a weak dependence on $B_\perp$, as shown in Extended Data Figure 4, consistent with the Hofstadter spectrum in Fig. 3c. 

\begin{figure}
\includegraphics[width=\textwidth]{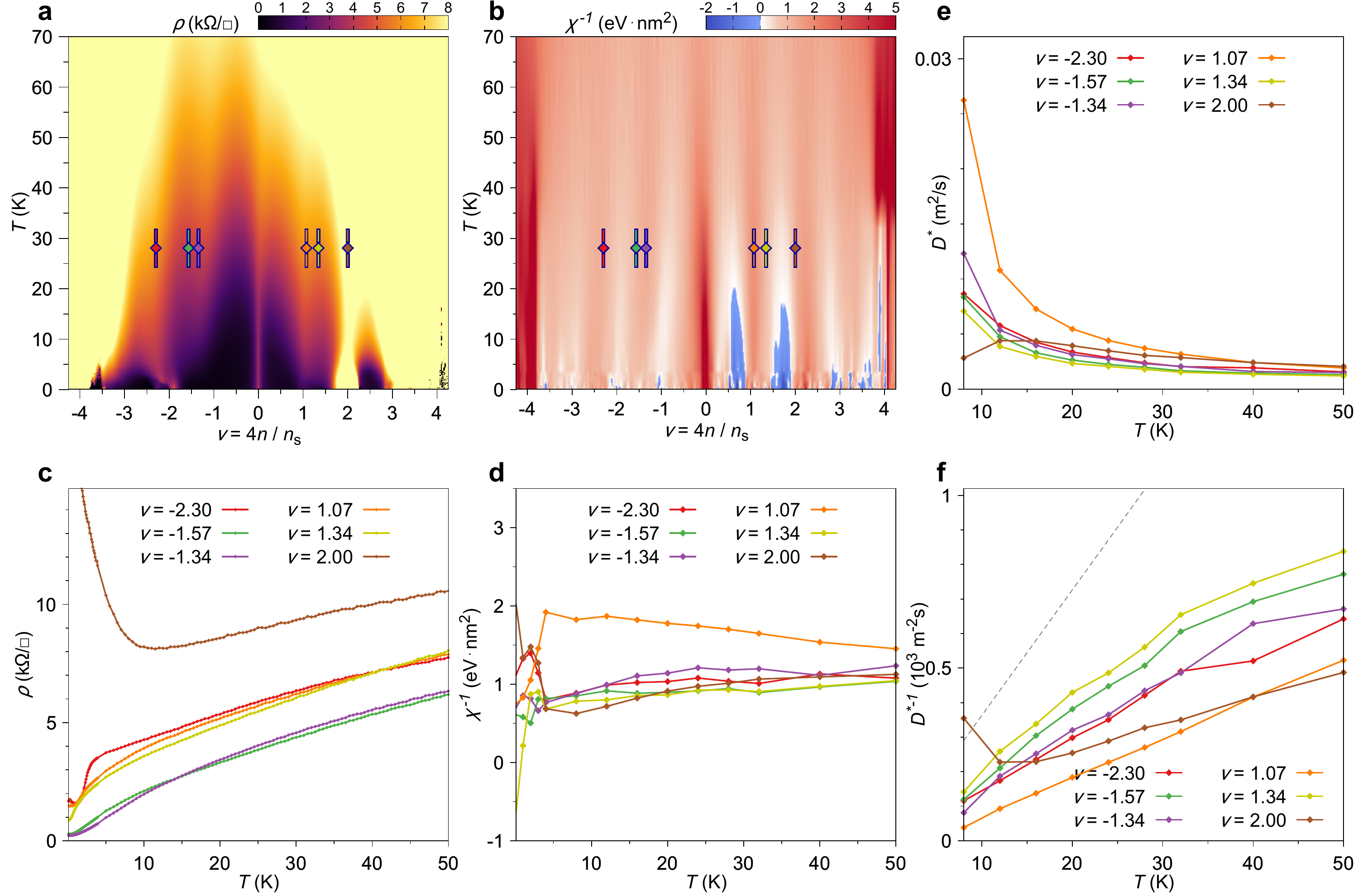}
\caption{\small Resistivity, electronic compressibility, and diffusivity of MATBG in the strange metal regime. (a) Resistivity and (b) inverse electronic compressibility $\chi^{-1}=d\mu/dn$ of MATBG versus $\nu$ and temperature. Colour marks show the position of $\nu$ where the line-cuts are taken in (c-f). (c) Linear resistivity-temperature behaviour across a range of densities around the correlated states, with only a weak dependence of slope on $\nu$. (d) Line-cuts of $\chi^{-1}$ do not show significant dependence on $T$. (e) Effective diffusivity $D^{*}=\chi^{-1}/e^2(\rho-\rho_0)$, where $\rho_0$ is obtained by fitting the linear $T$ range and extrapolating to $T=0$. (f) $1/D^{*}$ shows linear trend as a function of $T$. The gray dash line denotes a diffusivity bound $D_\mathrm{bound}(T)=\hbar v_F^2/(k_B T)$, where we used a Fermi velocity of $v_F=\SI{6e4}{\meter\per\second}$.}
\end{figure}

In correlated metals with multiple bands near the Fermi energy, the atomic Hund's coupling is known to play an important role in their many-body physics, including the strange metal regime\cite{georges_strong_2013}. In MATBG, recent experiments have reported evidence for strange metal behaviour\cite{cao_strange_2020}, manifested as resistivity linear with temperature, from very low $T$ up to $T$ above the Fermi energy. As shown in Fig. 4a and c, the resistivity in our MATBG sample is largely linear with $T$ over a range of densities around the correlated states, and with a slope that is weakly dependent on $n$, consistent with recent works\cite{cao_strange_2020, polshyn_large_2019}. The resistivity keeps increasing with $T$ without any sign of saturation up to \SI{50}{\kelvin}, suggesting non-Fermi liquid transport in this system\cite{cao_strange_2020}. It has been hypothesized that the strange metal behaviour can be universally described by a `Planckian' scattering rate bound $\Gamma\sim k_BT/\hbar$ in the framework of incoherent non-quasiparticle transport \cite{zaanen_why_2004, bruin_similarity_2013}. However, the construction of a microscopic picture for this bound is still in progress\cite{chowdhury_translationally_2018, patel_theory_2019}.

An appropriate framework to investigate the strange metal regime, regardless of the existence or absence of quasiparticles, is the Nernst-Einstein relation, which connects the resistivity $\rho$, compressibility $\chi$, and charge diffusivity $D$ of a generic conductor by $\rho^{-1}=e^2\chi D$. A linear in $T$ resistivity could thus originate from: (i) $\chi^{-1}\propto T$, which could come from thermodynamic contributions when $k_{B}T\gtrsim W$\cite{hartnoll_theory_2015,perepelitsky_transport_2016}; from (ii) $D^{-1}\propto T$, which would represent a linear scattering rate; or (iii) from a more complex combined $T$ dependence of both. Differentiating between these possibilities could help constrain theoretical models for strange metal behaviour\cite{hartnoll_theory_2015,pakhira_absence_2015,kokalj_bad-metallic_2017}. However, to the best of our knowledge, there are no reported measurements of the electronic compressibility or charge diffusivity for any correlated materials in the strange metal regime, and only recent experiments have begun to explore this physics in the very high temperature regime in ultracold atom systems\cite{brown_bad_2019}.

Our combined resistivity and compressibility measurements allow us to extract the charge diffusivity of MATBG. Figure 4b shows the inverse compressibility $\chi^{-1}$ as a function of $\nu$ and $T$. While for $T<\SI{20}{\kelvin}$ $\chi^{-1}$ becomes negative before each integer filling factor, as discussed above, at higher $T$ it converges to a roughly constant value of order $\SI{1}{\electronvolt\nano\meter\squared}$ for any value of $\nu$. Figure 4d shows some representative traces of $\chi^{-1}$ vs $T$ for the same $\nu$ values as in Fig. 4c. The traces exhibit only a weak dependence on $T$, albeit for all these densities $\rho$ exhibits a prominent linear $T$ behaviour. Therefore, this suggests that the linear $\rho$-$T$ behaviour in MATBG is mainly due to a $T$-dependent charge diffusivity $D$. Figure 4e shows the $T$-dependence of the extracted effective diffusivity $D^{*}=\chi^{-1}/e^2(\rho-\rho_0)$, where $\rho_0$ is the residual resistivity extrapolated at zero temperature, and Fig. 4f shows its inverse $1/D^{*}$. These quantities indeed do appear to roughly follow a $\sim T^{-1}$ and a $\sim T$ trend, respectively. Our observations therefore indicate that the strange metal transport regime in MATBG is consistent with a scattering rate linear in $T$. Note that these arguments do not apply to regions with negative electronic compressibility and thus negative $D^{*}$, as the interpretation of diffusivity in this case needs to be modified\cite{efros_negative_2008} (see also Supplementary Information for relevant data and discussion). Interestingly, we find the extracted diffusivity $D^*(T)$ at all these fillings to be within about a factor of 2 from a diffusivity bound $D_\mathrm{bound}=\hbar v_F^2/(k_B T)$ proposed for incoherent metals\cite{hartnoll_theory_2015}, using $v_F=\SI{6e4}{\meter\per\second}$ estimated above at low temperatures. While this bound is known to be violated in the low-temperature region in a large-$U$ system\cite{pakhira_absence_2015}, this is not at odds with our observations if MATBG is in the intermediate $U$ regime ($U/W\sim1$), as suggested by the range of $U$ that qualitatively reproduce our experimental data, as well as other recent experiments.\cite{xie_spectroscopic_2019,zondiner_cascade_2020,wong_cascade_2020}

\bibliographystyle{Nature}
\bibliography{references} 

\section*{Acknowledgements}
The authors thank Antoine Georges, Francisco Guinea, Shahal Ilani, Aharon Kapitulnik, Leonid Levitov, Louis Taillefer, Senthil Todadri, Ashvin Vishwanath, Amir Yacoby, Denis Bandurin, Sergio de la Barrera, Clement Collignon, Ali Fahimniya, Daniel Rodan-Legrain, Yonglong Xie, and Kenji Yasuda for fruitful discussions.

\section*{Methods}

\subsection{Sample Fabrication}
The multilayer heterostructure consists of one sheet of monolayer graphene (MLG) and twisted bilayer graphene (TBG) twisted at a small angle $\theta\sim\SI{1.1}{\degree}$, separated by a thin ($\sim\SI{1}{nm}$) h-BN layer. This sandwich is encapsulated by two h-BN flakes. All flakes were first exfoliated on SiO\textsubscript{2}/Si substrates, and subsequently analyzed with optical microscopy and atomic force microscopy to determine their thicknesses and quality. The multilayer heterostructure was fabricated by a modified polymer-based dry pick-up technique, where a layer of poly(bisphenol A carbonate)(PC)/polydimethylsiloxane(PDMS) on a glass slide fixed on the micro-positioning stage was used to sequentially pick up the flakes. The order of the pick-up was h-BN-MLG-h-BN(\SI{1}{nm})-MLG-MLG, where the last two MLG sheets were laser-cut from one MLG flake (see Supplementary Information) and twisted by an angle $\sim\SI{1.1}{\degree}$. All h-BN layers were picked up at \SI{90}{\celsius}, while the MLG layers were picked up at room temperature. The h-BN-MLG-h-BN(\SI{1}{nm})-MLG-MLG heterostructure was then released on the pre-stacked h-BN-Pd/Au back gate at \SI{175}{\celsius}. Hall-bar geometry for transport measurements was defined with electron beam lithography and reactive ion etching for each of the MLG and MLG-MLG layers. The top gate and electrical edge-contacts were patterned with electron beam lithography and thermal evaporation of Cr/Au. 

\subsection{Measurement Setup}
Electronic transport measurements were performed in a dilution refrigerator with a superconducting magnet, with a base electronic temperature of \SI{70}{\milli\kelvin}. Current through the sample, amplified by \SI{1e7}{\volt\per\ampere}, and the four-probe voltage, amplified by $1000$, were measured with SR-830 lock-in amplifiers synchronized at the same frequency between \SIrange{1}{20}{\hertz}. Current excitation of $\SI{1}{\nano\ampere}$ or voltage excitation of $\SI{50}{\micro\volt}$ to $\SI{100}{\micro\volt}$ was used for resistance measurements. We measured both MLG and MATBG layers simultaneously for accurate comparison.

\subsection{Maxwell's Relations}
Using Maxwell's relations between thermodynamic variables, we can obtain information about various thermodynamic quantities by taking different derivatives of the chemical potential. The free energy of the system per unit area in the presence of a magnetization can be written as $g=u-Ts+M_{\parallel}B_{\parallel}$, where $u, M, s$ are the internal energy, magnetization, and entropy per area respectively. $u$ and $g$ satisfy
\begin{align}
    du &= Tds + B_{\parallel}dM_{\parallel} + \mu d\nu,\\
    dg &= -sdT - M_{\parallel}dB_{\parallel} + \mu d\nu.
\end{align}

By taking the second derivative of $g$ with respective to $(\nu, B)$ in different orders, we can obtain the following Maxwell's relationship,
\begin{align}
\left(\frac{\partial M_{\parallel}}{\partial \nu}\right)_{T,B_{\parallel}} &= -\left(\frac{\partial\mu}{\partial B_{\parallel}}\right)_{T,\nu}.
\end{align}

Therefore, we can integrate from the $B_{\parallel}$-derivative of $\mu$ to obtain the change in $M_{\parallel}$ as a function of density $\nu$,
\begin{align}
M_{\parallel} &= M_{\parallel}(\nu=0) - \int_0^\nu \left(\frac{\partial\mu}{\partial B_{\parallel}}\right)_{T,n} dn
\end{align}

The extracted $\partial M_{\parallel}/\partial \nu$ and $M_{\parallel}$ versus $\nu$ are shown in Extended Data Figure 3. We extract the uncertainty (\SI{95}{\percent} confidence interval) of $\partial M_{\parallel}/\partial \nu$ from fitting of $\mu$ with $B_\parallel$, and propagate through the integration to obtain uncertainty in $M_{\parallel}$.

\subsection{Thermal activation gap analysis}
Thermal activation gap analysis was performed based on the Arrhenius formula $R \sim \exp(-\Delta/2k_{B}T)$, where $k_B$ is the Boltzmann constant and $\Delta$ is the gap size. A temperature-dependent background was removed from the raw resistance $R_{xx}$ of MATBG to avoid being affected by the linear $R_{xx}$-$T$ behaviour in MATBG.\cite{wu_chern_2020} The corrected quantity is denoted by $R^{*}_{MATBG}$ and shown in Extended Data Fig. 2a-b. By fitting the gaps as a function of the in-plane magnetic field $B_{\parallel}$ to $\Delta=g\mu_BB_\parallel$, where $\mu_B$ is the Bohr magneton, we find effective transport $g$-factors of $\sim1.31$ for the $\nu=+2$ state and $\sim0.57$ for the $\nu=+1$ state, as shown in Extended Data Fig. 2c.

\clearpage
\section*{Extended Data Figures}

\renewcommand{\figurename}{Extended Data Figure}
\setcounter{figure}{0}

\begin{figure}[!ht]
\includegraphics[width=\textwidth]{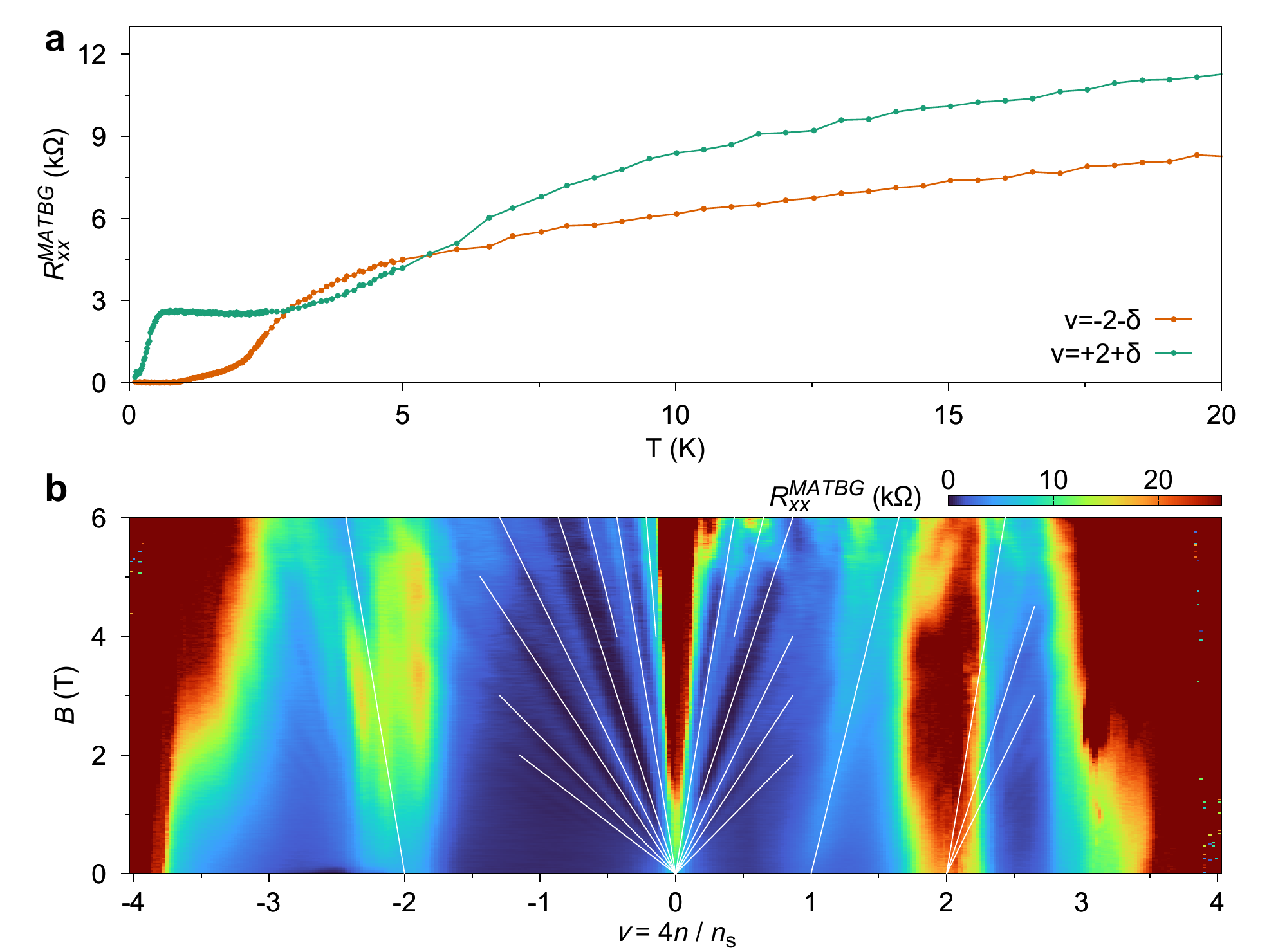}
\caption{\small Superconductivity and Landau fan diagram of MATBG. (a) Superconducting curves for $\nu=-2-\delta$ and $+2+\delta$ domes of MATBG. (b) Landau fan diagram of MATBG at $\SI{1}{\kelvin}$. The CNP shows the main sequence $\nu_{LL}=\pm4, \pm8, \ldots$ and broken symmetry states $\nu_{LL}=-1, \pm2, \pm3$. There are fans from $\nu=\pm2$, where the sequence $\nu_{LL}=+2, +4, +6$ and $\nu_{LL} =-2$ are seen, respectively. We also find transport evidence of a correlated Chern gap with Chern number $C=3$ from $\nu=+1$.}
\end{figure}

\clearpage

\begin{figure}
\includegraphics[width=\textwidth]{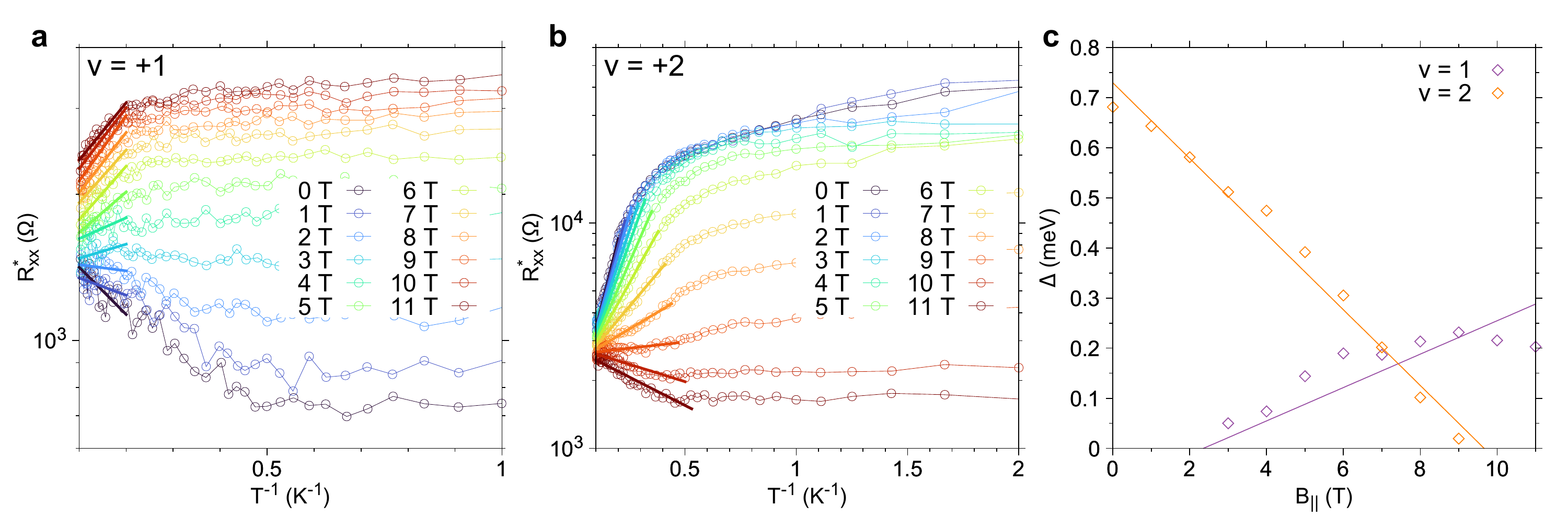}
\caption{\small Thermal activation gap analysis and $g$-factors of the correlated states. (a)-(b) Fitting of temperature-dependent resistance using the Arrhenius formula $R^{*}\sim \exp(-\Delta/2k_{B}T)$ at $\nu=+1,+2$, respectively, for in-plane magnetic fields $B_{\parallel}=\SI{0}{T}-\SI{11}{T}$. $R^*$ is the background-removed resistance of MATBG. (c) $B_{\parallel}$-dependence of the thermal activation gap $\Delta$. The extracted g-factors are $\sim0.57, 1.31$ for the $\nu=+1,+2$ states, respectively.}
\end{figure}

\clearpage

\begin{figure}
\includegraphics[width=\textwidth]{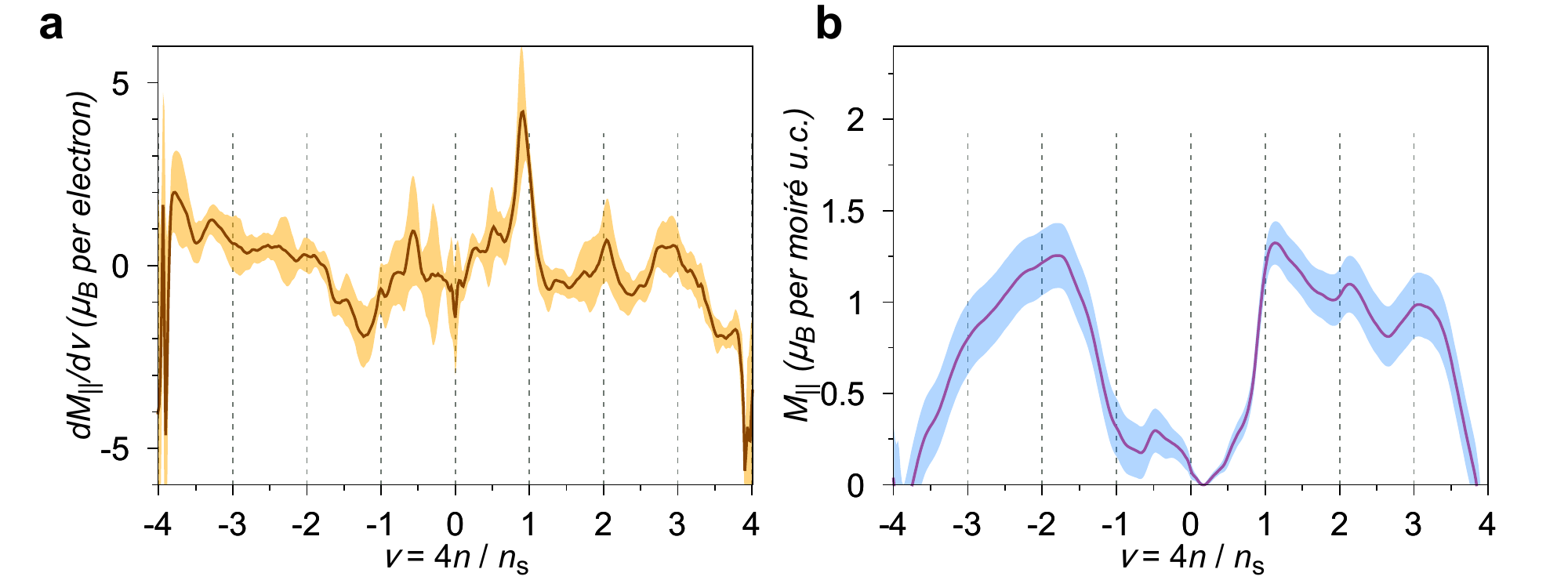}
\caption{\small In-plane magnetization of MATBG. (a) $\left(\frac{\partial M_{\parallel}}{\partial\nu}\right)_{B_{\parallel},T} = -\left(\frac{\partial\mu}{\partial B_{\parallel}}\right)_{\nu,T}$ versus $\nu$. Peaks are visible near $\nu=\pm1$. $T=\SI{4}{\kelvin}$. (b) Magnetization $M_{\parallel}$ from integrating the curve in (a). $M_{\parallel}$ persists near all filling factors $\nu=\pm1,\pm2,\pm3$. The error bands correspond to a confidence level of \SI{95}{\percent}.}
\end{figure}

\clearpage

\begin{figure}
\includegraphics[width=\textwidth]{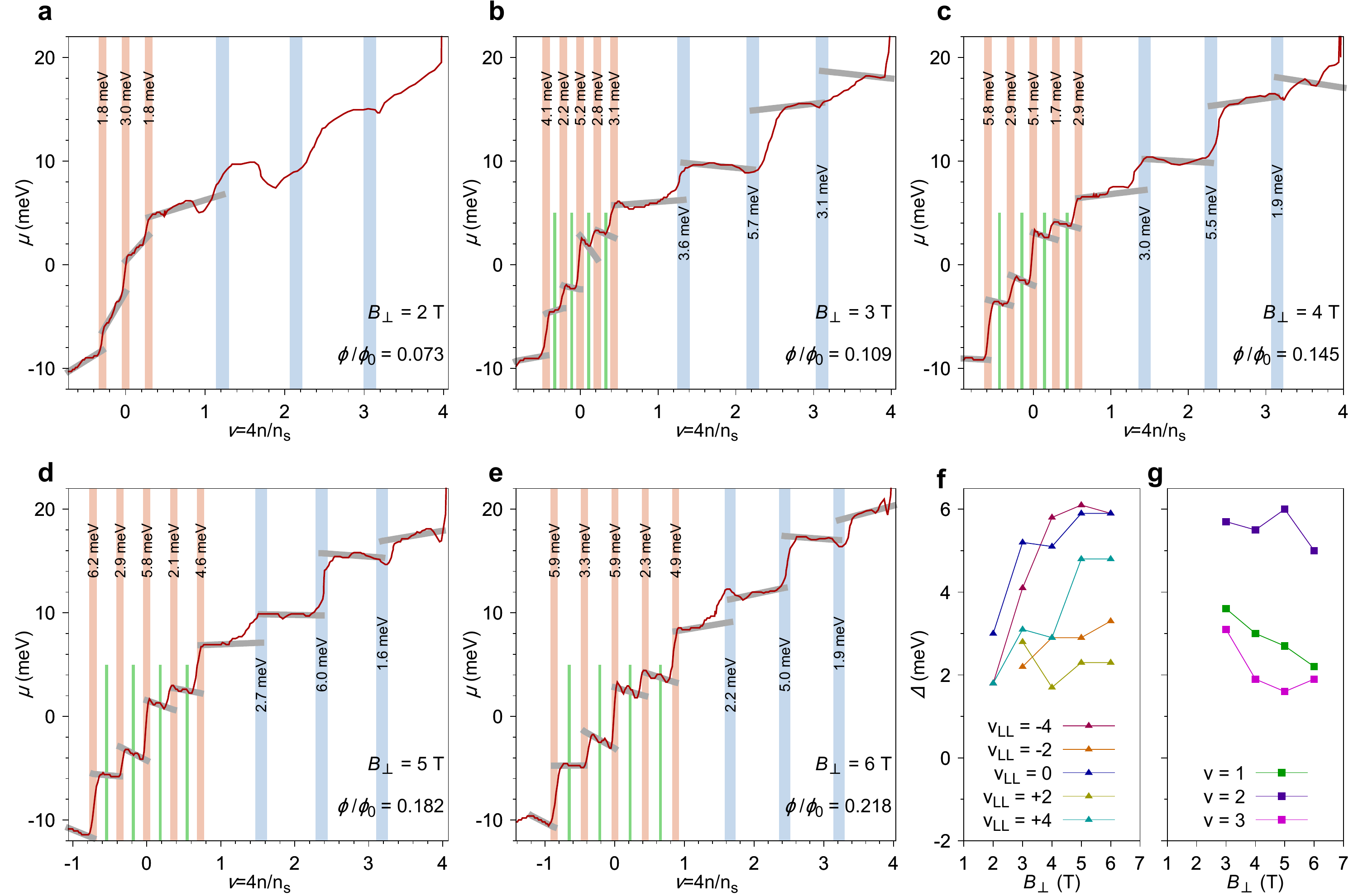}
\caption{\small Magnetic field $B_{\perp}$ dependence of Landau level and Chern gaps. (a)-(e) Gap extraction from the chemical potential curves at $B=\SI{2}{T}-\SI{6}{T}$. (f-g) Magnetic field dependence of (f) the Landau level gaps and (g) the correlated Chern gaps. While the $\nu_{LL}=\pm4, 0$ Landau level gaps have increasing trend with $B_{\perp}$, the $\nu_{LL}=\pm2$ gaps show relatively weak dependence. Similarly, the correlated Chern gaps also exhibit weak dependence on $B_{\perp}.$}
\end{figure}

\clearpage

\begin{figure}
\includegraphics[width=\textwidth]{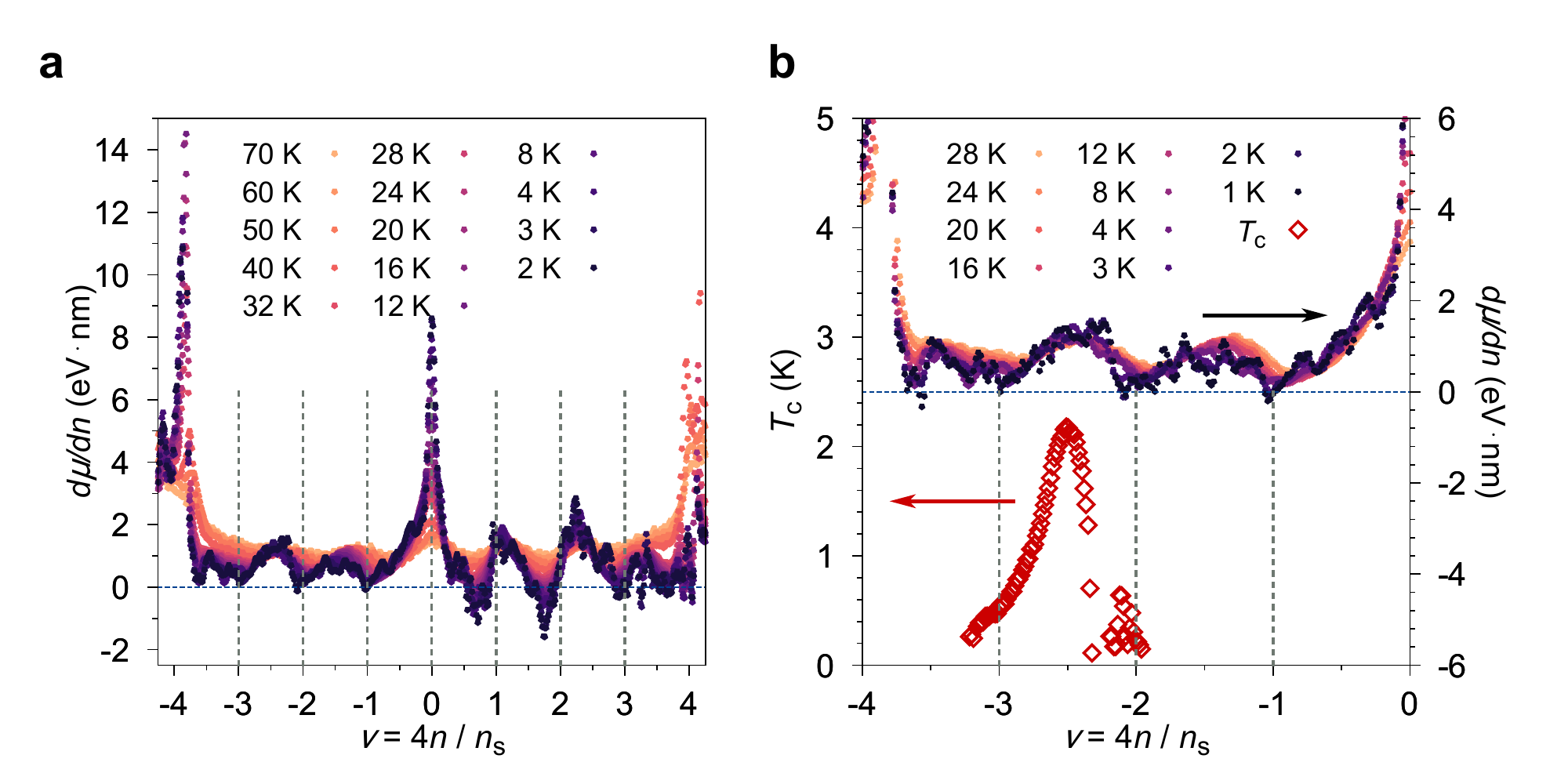}
\caption{\small Overlaying the inverse compressibility and the superconducting dome. (a) Temperature dependence of inverse compressibility $d\mu/dn$ for $T=\SI{2}{K}\sim\SI{70}{K}$ at $B=\SI{0}{T}$. Negative compressibility near $\nu=+1,+2$ persists up to $T\sim\SI{20}{K}$. (b) Comparison between $d\mu/dn$ and superconducting $T_{c}$ dome (red points, 20\% normal-state resistance) near $\nu=-2-\delta$. The $T_{c}$ dome occurs near maximum $d\mu/dn$, which is unexpected within weak coupling BCS-type mechanism for the superconductivity.}
\end{figure}

\clearpage
\newpage



\section*{Supplementary material}

# Supplementary Information

## CONTENTS

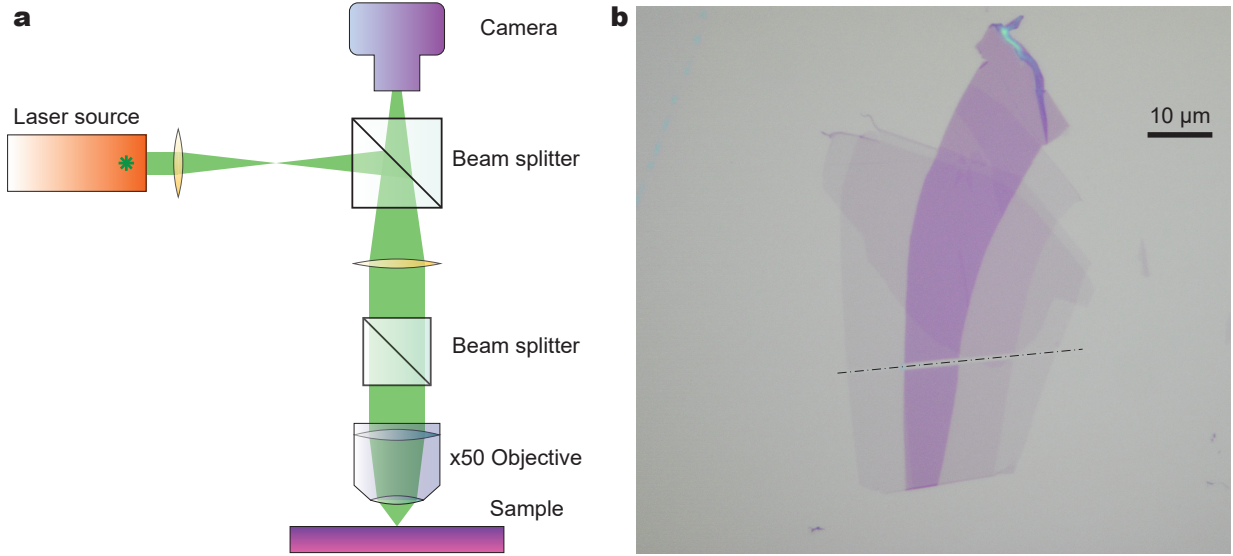

FIG. S1. Laser-cutting of graphene. (a) Setup for laser-cutting integrated into a standard microscope. (b) Example of graphene cut by the laser, showing a clean-cut across multiple layers.

# I.  LASER CUT & STACK

Conventionally, TBG is prepared by tearing a single-crystal graphene flake and then stacking the two parts together, or namely the 'tear & stack' method[1, 2]. The tearing relies on the stiction between the top h-BN flake and the part of graphene it covers. However, this method has several drawbacks. Firstly, it has a limitation on the thickness of the top h-BN. We find that ultra thin h-BN ($\leq 2\,\mathrm{nm}$) is not suitable for tearing graphene, because the stiction is very weak. Secondly, the graphene near the torn edge is typically not clean and can roll up or develop cracks. Thirdly, we suspect that the tearing can introduce excess strain in the sample, which increases the twist angle disorder.

To overcome these issues and improve the sample quality, we developed a new method of cutting graphene using a focused laser beam, and we call it the 'laser-cut & stack' method. The schematic of the setup is illustrated in Fig. S1a. The laser source and the related optics are all integrated into the standard dry-transfer setup. The main addition is a laser source and a beam splitter in the light path before the camera. For cutting graphene, we use a supercontinuum laser with output filtered to visible range ($400\,\mathrm{nm}$ to $700\,\mathrm{nm}$) with tunable output power. The supercontinuum laser has a pulse width on the order of $\sim 10\,\mathrm{ps}$ and a repetition rate of $40\,\mathrm{MHz}$. The laser is focused into a submicron spot using a standard objective (50x, NA=0.75), and its light path is confocal with the normal imaging path through the microscope. The graphene can be reliably cut while the silicon

oxide subtrate remains intact in the output power range of $60\,\mathrm{mW}$ to $150\,\mathrm{mW}$ (measured before the beam splitter), which translates to an ablation threshold of about $130\,\mathrm{mJ\,cm^{-2}}$, assuming an effective spot diameter of $1\,\mu\mathrm{m}$ and transmission of $70\,\%$ inside the microscope, consistent with typical values widely reported in the literature, see e.g. [3–6]. The pulse width and wavelength of the laser appears to be unimportant for cutting graphene as long as it is less than nanosecond scale. We find however that shorter pulse width tends to create cleaner cut edges, which is expected as thermal effects are reduced at shorter time scales.

Fig. S1b shows an example of the laser-cut graphene. We find that few-layer graphene and graphite up to tens of nanometer thick can be easily cut with the same power, as thicker graphene/graphite also has higher absorption. The laser ablation of graphene leaves very little residue on its surface as the products are likely volatile, as can be seen from the optical image. The rest of the sample fabrication proceeds in the same way as the 'tear & stack' method.

## II. CHEMICAL POTENTIAL EXTRACTION

### A. Conversion Formulae

In the presence of another layer, the electric field from the top and back gates do not fully penetrate to the layer of interest due to screening. This principle allows us to measure the chemical potential as a function of carrier density in the double-layer heterostructure device. Here, we will derive the relations between the experimentally controlled quantities, such as gate voltages, and the physical quantities of interest, such as chemical potential and carrier densities. When there are two electronic systems in equilibrium, their electrochemical potential must be equal. We consider the case where the MATBG and MLG layers are shorted to ground while we apply voltage to the top and back gates. Figure 1b (main text) shows the increase of electric potential from MATBG to MLG ($V_0$), from MATBG to back gate ($V_1$), and from MLG to top gate ($V_2$). Balancing the electrochemical potential, the following relations are obtained

$$eV_1 = eV_{bg} - \mu_{\mathrm{MATBG}}(n_{\mathrm{MATBG}}) \tag{1}$$

$$eV_2 = eV_{tg} - \mu_{\mathrm{MLG}}(n_{\mathrm{MLG}}) \tag{2}$$

$$eV_0 = \mu_{\mathrm{MLG}}(n_{\mathrm{MLG}}) - \mu_{\mathrm{MATBG}}(n_{\mathrm{MATBG}}) \tag{3}$$

The densities in terms of electric potentials are given by the electrostatics equations

$$en_{\text{MATBG}} = C_{bg}V_1 + C_iV_0 \tag{4}$$

$$en_{\text{MLG}} = C_{tg}V_2 - C_iV_0 \tag{5}$$

where $C_{bg}, C_{tg}, C_i$ are the geometric capacitances per unit area of the bottom, top, and interlayer h-BN dielectric layers, respectively. Note that the sign conventions in all the equations are chosen such that the arrow shown in the diagram denotes the positive direction.

Combining these five equations, the following two master equations are obtained that relate $(V_{tg}, V_{bg}, C_{tg}, C_{bg}, C_i)$ and $(n_{\text{MATBG}}, n_{\text{MLG}}, \mu_{\text{MATBG}}, \mu_{\text{MLG}})$:

$$\begin{cases} en_{\text{MATBG}} & = C_{bg}(V_{bg} - \frac{\mu_{\text{MATBG}}(n_{\text{MATBG}})}{e}) + C_i(\frac{\mu_{\text{MLG}}(n_{\text{MLG}})}{e} - \frac{\mu_{\text{MATBG}}(n_{\text{MATBG}})}{e}) \\ en_{\text{MLG}} & = C_{tg}(V_{tg} - \frac{\mu_{\text{MLG}}(n_{\text{MLG}})}{e}) - C_i(\frac{\mu_{\text{MLG}}(n_{\text{MLG}})}{e} - \frac{\mu_{\text{MATBG}}(n_{\text{MATBG}})}{e}) \end{cases} \tag{6}$$

Two simple cases are obtained when the carrier density of one layer is kept constant. At MATBG charge neutrality point, where $n_{\text{MATBG}} = 0$ and $\mu_{\text{MATBG}} = 0$, we can obtain the following:

$$\begin{cases} n_{\text{MLG}} & = \frac{C_{tg}V_{tg}}{e} + \frac{(C_{tg}+C_i)C_{bg}V_{bg}}{eC_i} \\ \mu_{\text{MLG}} & = -\frac{eC_{bg}V_{bg}}{C_i} \end{cases} \tag{7}$$

This means that the charge-neutrality feature of MATBG in the $V_{tg}$ and $V_{bg}$ space can be directly converted to $\mu_{\text{MLG}}(n_{\text{MLG}})$, the chemical potential of MLG as a function of carrier density.

Similarly, if one tracks the MLG charge neutrality point, where $n_{\text{MLG}} = 0$ and $\mu_{\text{MLG}} = 0$, the equations for MATBG are given by:

$$\begin{cases} n_{\text{MATBG}} & = \frac{C_{bg}V_{bg}}{e} + \frac{(C_{bg}+C_i)C_{tg}V_{tg}}{eC_i} \\ \mu_{\text{MATBG}} & = -\frac{eC_{tg}V_{tg}}{C_i} \end{cases} \tag{8}$$

### B. Extraction with MLG Charge Neutrality

As discussed in the main text, the MATBG chemical potential can be extracted from the $V_{tg}$-$V_{bg}$ maps of $R_{xx}^{\text{MLG}}$ by tracking the MLG charge neutrality (Dirac point). A self-consistent extraction procedure is performed on the $V_{tg} - V_{bg}$ resistance maps of both MLG and MATBG to solve the nonlinear equations Eq. 6, as detailed in the following. The capacitances are extracted using alternative gating configurations, as detailed in Section II C.

1. Extract $\mu_{\text{MLG}}(n_{\text{MLG}})$ from the charge neutrality peak ($\mu_{\text{MATBG}} = n_{\text{MATBG}}=0$) in the $R_{xx}^{\text{MATBG}}$ map, using Eq. 7.

2. Initialize $\mu_{\text{MATBG}}(n_{\text{MATBG}})$ with an arbitrary (e.g. linear) function.

3. The $V_{tg} - V_{bg}$ resistance maps are converted to $n_{\text{MATBG}} - n_{\text{MLG}}$ maps using Eq. 6, and the functions $\mu_{\text{MLG}}(n_{\text{MLG}})$ and $\mu_{\text{MATBG}}(n_{\text{MATBG}})$.

4. We extract the resistance peak position (corresponding to the charge neutrality point of MLG) in $n_{\text{MLG}}$, for each $n_{\text{MATBG}}$ (see Sec. VII for extraction details). If the peak position $n_{\text{peak}}$ is close enough to 0 for all $n_{\text{MATBG}}$, indicating that $\mu_{\text{MATBG}}$ has converged, we stop the iteration.

5. The new $\mu_{\text{MATBG}}$ is calculated by $\mu_{\text{MATBG}}(n_{\text{MATBG}}) - h \cdot en_{\text{peak}}/C_i$, where $h$ is a numerical factor controlling the convergence. In the next iteration, the resistance peak positions shall be closer to zero.

6. Repeat from step 3.

## C. Using Alternative Gating Configuration

We can similarly extract the chemical potential by applying voltage $V_b$ between the MLG and MATBG layers and applying only one of the two gate voltages, as was performed in Ref. [7, 8]. Combining these results with the $V_{tg}$-$V_{bg}$ configuration, we can accurately determine the values of the geometric capacitance $C_{bg}, C_{tg}, C_i$ and the Fermi velocity of MLG, $v_F$. With a bias voltage $V_b$ on MLG (MATBG is kept at ground), the formulae in Eq. 6 are modified to

$$
\begin{cases}
en_{\text{MATBG}} & = C_{bg}(V_{bg} - \frac{\mu_{\text{MATBG}}(n_{\text{MATBG}})}{e}) + C_i(\frac{\mu_{\text{MLG}}(n_{\text{MLG}})}{e} - \frac{\mu_{\text{MATBG}}(n_{\text{MATBG}})}{e} + V_b) \\
en_{\text{MLG}} & = C_{tg}(V_{tg} - V_b - \frac{\mu_{\text{MLG}}(n_{\text{MLG}})}{e}) - C_i(\frac{\mu_{\text{MLG}}(n_{\text{MLG}})}{e} - \frac{\mu_{\text{MATBG}}(n_{\text{MATBG}})}{e} + V_b)
\end{cases}
\tag{9}
$$

First we wish to obtain the ratio between $C_{tg}, C_{bg}$ and $C_i$. At MATBG charge neutrality, Eq. 9 becomes

$$
\begin{cases}
\mu_{\text{MLG}} & = eV_b - \frac{eC_{bg}}{C_i}V_{bg} \\
n_{\text{MLG}} & = \frac{C_{tg}}{e}V_{tg} + \frac{C_{tg}+C_i}{eC_i}C_{bg}V_{bg}
\end{cases}
\tag{10}
$$

Interestingly, if $V_{bg}$ is also set to zero, $V_{tg}$ and $V_b$ are directly proportional to $n_{\text{MLG}}$ and $\mu_{\text{MLG}}$ respectively, as shown in Fig. S2a for example. If we add $V_{bg}$ now, the shift of the feature in

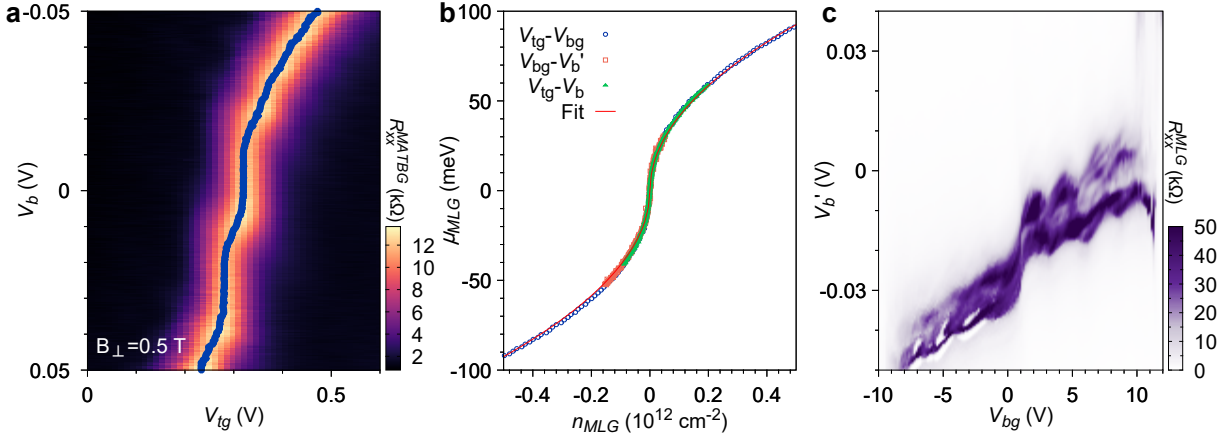

FIG. S2. Measurement of the chemical potential using other gating configurations. (a) The chemical potential of MLG can be directly obtained from the $V_{tg} - V_b$ map of $R_{xx}^{\mathrm{MATBG}}$. Here the perpendicular magnetic field is 0.5 T, and steps in position of the resistance peak can be clearly identified, which correspond to the Landau level gaps of MLG. (b) The chemical potential extracted from all possible gating configuration can be fit with a Fermi velocity of $1.12 \times 10^6 \, \mathrm{m \, s^{-1}}$. (c) The $R_{xx}^{\mathrm{MLG}}$ map in the $V_{bg}$-$V_b'$ space provides similar information as Fig. 2a, from which we can obtain the same information on the chemical potential.

the $V_{tg} - V_b$ map directly gives us the capacitance ratio $C_{bg}/C_i$ and $(C_{tg} + C_i)C_{bg}/C_i$ with a high precision. From this shift we obtain $C_{bg}/C_i = 0.0263$ and $C_{tg}/C_i = 0.109$ respectively. Note that this determination does not depend on any assumption about the chemical potential of either layer.

Next, we pinpoint the absolute value of the capacitances by fitting the Landau fan diagram in strong perpendicular magnetic fields, which is shown in Extended Data Figure 1. The procedure of fitting the Landau fan in TBG is detailed in previous works[1, 9, 10]. The fitting gives us the capacitances $C_{bg} = 4.60 \times 10^{-4} \, \mathrm{F \, m^{-1}}$, $C_{tg} = 1.90 \times 10^{-3} \, \mathrm{F \, m^{-1}}$, and $C_i = 1.75 \times 10^{-2} \, \mathrm{F \, m^{-1}}$. The twist angle is determined to be $(1.07 \pm 0.03)°$.

With these values, we can determine the Fermi velocity in the MLG, by fitting the measured chemical potential and density in MLG at MATBG charge neutrality from Eq. 10 to the formula $\mu_{\mathrm{MLG}} = \hbar v_F \sqrt{\pi |n_{\mathrm{MLG}}|} \mathrm{sgn}(n_{\mathrm{MLG}})$. Fig. S2b shows the extracted $n_{\mathrm{MLG}}$ and $\mu_{\mathrm{MLG}}$ from measurements performed in different gating configurations with $V_{bg}$, $V_{tg}$, and $V_b$. All data points fit well to a Fermi velocity of $v_F = 1.12 \times 10^6 \, \mathrm{m \, s^{-1}}$.

Similarly to Eq. 10, we can use the configuration $V_{bg} - V_b'$ to directly measure $\mu_{\mathrm{MATBG}}$ and $n_{\mathrm{MATBG}}$. $-V_b'$ is the bias voltage applied to MATBG while MLG is grounded (note the minus sign for consistency with previous discussion about $V_b$). $\mu_{\mathrm{MATBG}}$ and $n_{\mathrm{MATBG}}$ are proportional to $V_b'$ and $V_{bg}$ respectively if $V_{tg} = 0$. Fig. S2c shows $R_{xx}^{\mathrm{MLG}}$ measured in this configuration. The features

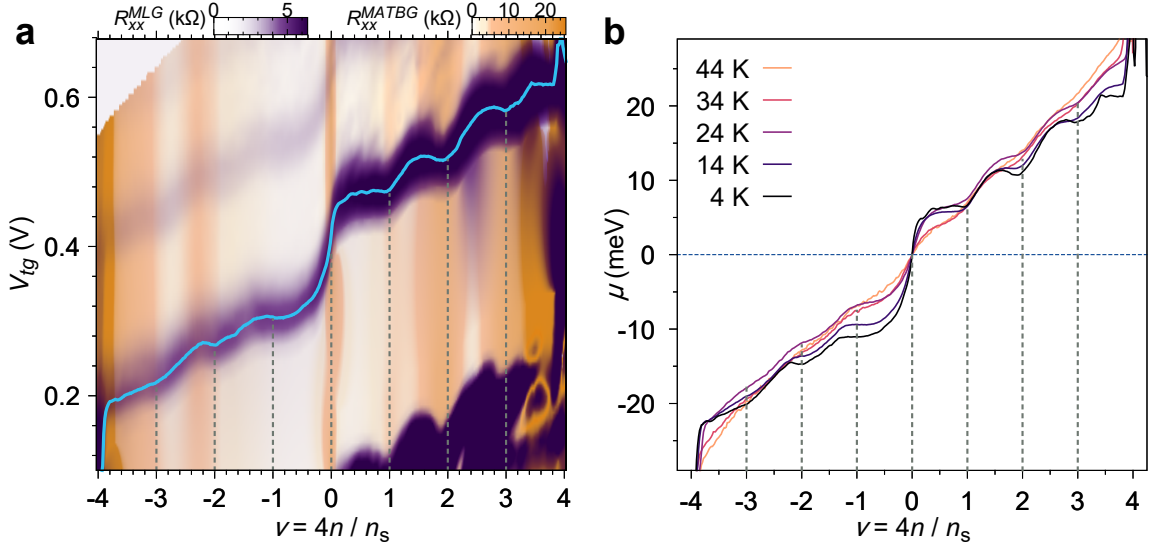

FIG. S3. Measurement of the chemical potential using a MLG Landau level. (a) Chemical potential of MATBG can also be extracted from the $N = 1$ Landau level transition of the MLG. (b) Temperature dependence of $\mu_{\mathrm{MATBG}}$ extracted using the MLG $N = 1$ Landau level.

are largely the same as what we identified with the default configuration $V_{tg} - V_{bg}$, as shown in Fig. 2a. We mainly used the $V_{tg} - V_{bg}$ configuration throughout the paper without using $V_b$ or $V_b'$ because we could then measure both $R_{xx}^{\mathrm{MATBG}}$ and $R_{xx}^{\mathrm{MLG}}$ simultaneously, and the measurement time is significantly reduced. The plots in Fig. S2 show that the different gating configurations indeed give equivalent information about the chemical potential.

## D. Extraction with MLG Landau Levels

In the presence of a small perpendicular magnetic field $B_\perp$, the Landau levels (LL) of MLG can be used as an alternative to tracking the Dirac point. Fig. S3a shows the full map of the the transport in MATBG and MLG in a perpendicular magnetic field $B_\perp = 0.7\,\mathrm{T}$, and the extracted chemical potential on the right axis. Fig. S3b shows the temperature dependence of the chemical potential extracted using the first Landau level. We find that the features extracted from Landau levels or the charge neutrality are largely consistent, and there is no obvious change of chemical potential in MATBG in this small perpendicular field, as opposed to the large field case where correlated Chern gaps are observed (see Sec. III).

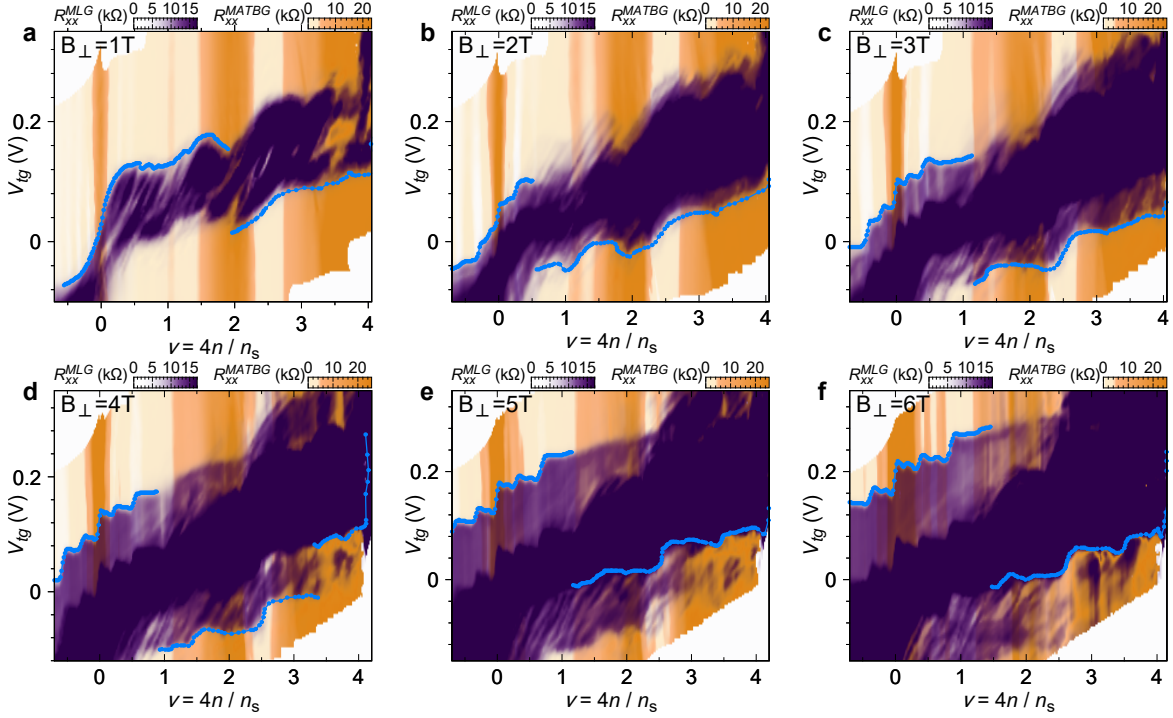

FIG. S4. Extraction of chemical potential in a perpendicular magnetic field. The data plotted are $R_{xx}^{\mathrm{MLG}}$ and $R_{xx}^{\mathrm{MATBG}}$ (see Fig. 2 caption for explanation). The blue dots denote the $\mu_{\mathrm{MATBG}}$ versus $\nu_{\mathrm{MATBG}}$ we extract for each field. Measurement temperature is 1 K.

## III. CHERN GAP ANALYSIS

To identify the correlated Chern gaps, we analyze the transport data measured in perpendicular magnetic fields from 1 T to 6 T, as shown in Fig. S4. Similar to Fig. 2 in the main text, here we also show the resistance of the MLG overlaid with the resistance of the MATBG. The perpendicular magnetic field induces a strong asymmetry on the MLG resistance, likely due to a contact resistance effect. To extract $\mu_{\mathrm{MATBG}}$ from these data (proportional to $V_{tg}$), we track different features for different ranges of $\nu_{\mathrm{MATBG}}$, as illustrated in Fig. S4. We subsequently stitch these different pieces together manually to obtain the full $\mu$-$\nu$ data as shown in Fig. 3a.

## IV. THEORETICAL MODELING

Our theoretical modeling follows the methodology of the mean-field model in [11] and [12]. We consider the total free energy of the system $G$ as a function of the total filling $\nu \in [-4, 4]$, which is the sum of the four filling flavours $\nu_1, \nu_2, \nu_3, \nu_4 \in [-1, 1]$. Each flavour has its own chemical

potential $\mu_i, i = 1, 2, 3, 4$, which is related to $\nu_i$ as

$$\nu_i = \int_0^{\mu_i} D(\epsilon) d\epsilon, \tag{11}$$

where $D(\epsilon)$ is the single-particle density of states (DOS). The free energy consists of a kinetic term and a Coulomb term

$$G = G_k + G_U, \tag{12}$$

$$G_k = \sum_i \int_0^{\mu_i} \epsilon D(\epsilon) d\epsilon, \tag{13}$$

$$G_U = U \sum_{i \neq j} \nu_i \nu_j - J \sum_i \nu_i^2. \tag{14}$$

In the last equation, the $U$ and $J$ captures the Coulomb repulsion and Coulomb exchange interaction respectively in the mean-field level of an extended Hubbard model.[13] The Coulomb repulsion $U$ has the largest contribution from on-site repulsion, which largely occur between electrons of different flavours due to the Pauli exclusion principle. On the contrary, the exchange term requires two interacting electrons to have the same flavour (for indistinguishability between two identical quantum particles upon exchanging), and therefore happens between different sites. Eq. 14 that contains the $U$ and $J$ terms can be rewritten as

$$G_U = (U + 2J) \sum_{i \neq j} \nu_i \nu_j - J \left( \sum_i \nu_i \right)^2 \tag{15}$$

$$= U' \sum_{i \neq j} \nu_i \nu_j - J \nu^2 \tag{16}$$

At equilibrium, the Gibbs free energy is minimized with respect to the four degrees of freedom $\nu_i$ (or equivalently, $\mu_i$). We numerically minimize $G$ as a function of these variables within the constraint that $\sum_i \nu_i = \nu$ and $\nu_i \in [-1, 1]$. The global chemical potential, which is determined by the chemical potential of the flavour(s) that is currently being filled, is calculated from its thermodynamic definition

$$\mu = \frac{\partial G}{\partial \nu}. \tag{17}$$

In Eq. 16, it is the effective repulsion $U' = U + 2J$ that controls the distribution of $\nu_i$. The second term in Eq. 16, on the other hand, creates a constant shift in the inverse compressibility,

$$\Delta G = -J \nu^2, \tag{18}$$

$$\Delta \mu = \frac{\partial G}{\partial \nu} = -2J\nu, \tag{19}$$

$$\Delta \chi^{-1} = \frac{\partial^2 G}{\partial \nu^2} = -2J. \tag{20}$$

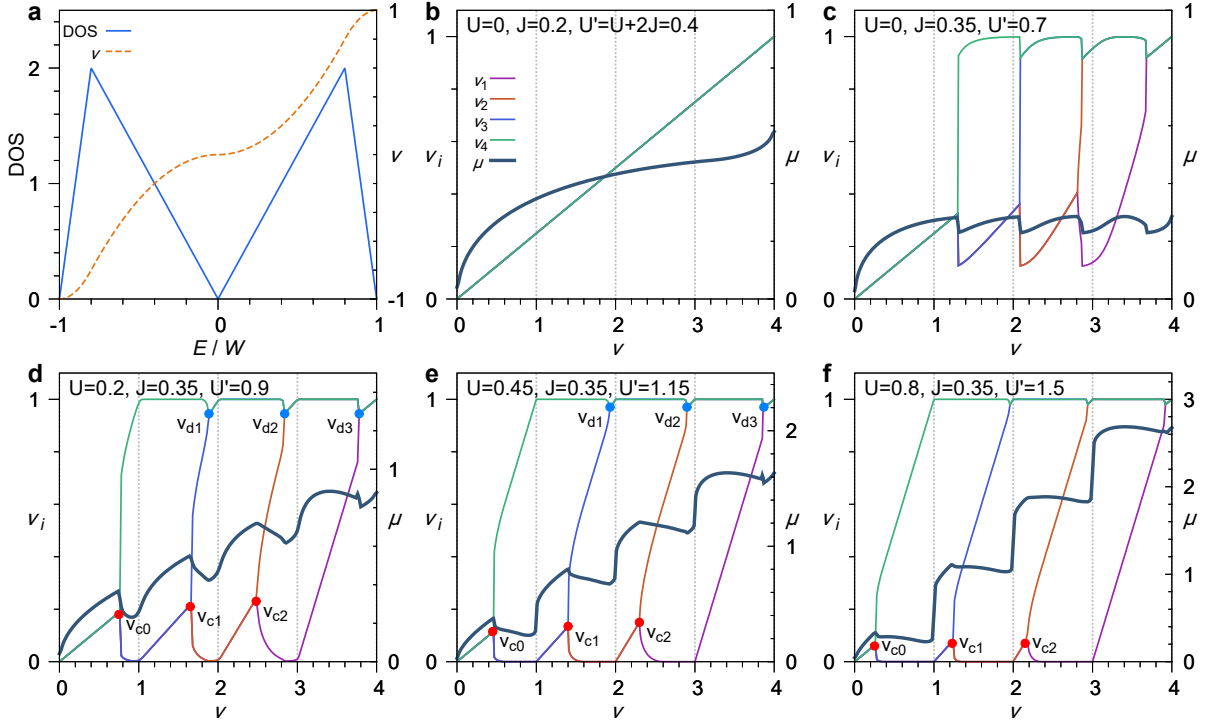

FIG. S5. Modeling flavour symmetry breaking in TBG at $B_\perp = 0$. (a) Triangular DOS we used to capture the essential features of the flat bands. The orange dashed line plotted on the right $y$-axis shows the filling $\nu$ versus energy. (b-f) Mean-field calculation of the filling per flavour $\nu_i, i = 1, 2, 3, 4$ and the chemical potential $\mu$ as a function of total filling $\nu$. The symmetry breaking is completely controlled by the parameter $U' = U + 2J$, while $J$ itself gives rise to negative compressibility. Note that in (c) where the interactions are barely enough to induce the phase transitions, the features in the chemical potential can occur away from the integers, which could explain the shift of features in the hole-doped side data.

When this negative contribution exceeds the other parts of the inverse compressibility coming from the single-particle DOS and the $U'$ term, the total $\chi^{-1} = d\mu/d\nu$ can be therefore negative. Our mean-field models for capturing the Coulomb physics in the presence and absence of a magnetic field are largely the same except for a few details, which we elaborate on below.

## A. Without Magnetic Field

The band structure of TBG without magnetic field has been well studied in the literature[14, 15]. The notable features of the single-particle DOS is the linear rise near the charge neutrality, and the divergence at the van Hove singularities (vHs). To significantly simplify the calculation while still correctly capturing the single-particle physics of the TBG bands, we approximate the DOS of

TBG as a triangular function, as shown in Fig. S5a. The energy axis and the density axis are now normalized in units of the bandwidth $W$ and full-filling density per flavour $n_s/4$, respectively, so $-1 < \nu_i, \mu_i < 1$. Furthermore, we also assume an exactly particle-hole symmetric spectrum with respect to the charge neutrality point, and for the simulation at zero magnetic field we limit both $\nu_i$ and $\mu_i$ to $[0, 1]$.

The flavour symmetry is evidently broken from the plot of $\nu_i, i = 1, 2, 3, 4$ versus the total $\nu = \sum_i \nu_i$, as shown in Fig. S5b-f. When the Coulomb interaction is not sufficient, the free energy is minimized at $\nu_i = \nu/4$, resulting in equal population of all flavours. As $U'$ is increased, we start to observe breaking of the flavour degeneracy (Fig. S5c). We notice that when $U'$ is barely large enough (Fig. S5c), the symmetry breaking is not complete and it occurs away from the integers, since the reduction in Coulomb repulsion at the integer fillings is not enough to compensate the kinetic penalty of breaking the symmetry. This regime may possibly explain the shifting of features we found on the hole-doped side in our experiment, if for example the hole-doped side bandwidth $W$ is larger than the electron-doped side.

Full flavour polarization is reached when the Stoner ferromagnetism criteria $U' \bar{D} > 1$ is reached, which occurs around $U' = 1$ since the average DOS $\bar{D}$ is on the order of 1[11]. Between densities $\nu = m$ and $\nu = m + 1$, $m = 0, 1, 2$, there is a symmetry-breaking phase transition at density $\nu_{cm}$. Before this transition density $\nu < \nu_{cm}$ there are $m$ fully filled flavours ($\nu_i = 1$) and $4 - m$ flavours with density $\nu_i = (\nu - m)/(4 - m)$, adding up to total density $\nu$. After this transition, there are $m$ fully filled flavours, one 'currently filling' flavour with density $\nu - m$, and $3 - m$ flavours with density zero $\nu_i = 0$ [11].

We find that the shape of the DOS has a small but noticeable effect on the symmetry breaking and the chemical potential, most notably right before $\nu$ reaches integer values larger than 1. This means that the chemical potential of each flavour has a tendency to get 'pinned' at the maximum of the DOS, so that the Coulomb energy can be minimized without raising the kinetic energy. In the case where the emulated "vHs" is far from the band edge, as shown in the case in Fig. S6b and c, there is another series of phase transition at a different density $\nu_{dm}, m = 1, 2, 3$. In the region $\nu_{dm} < \nu < m+1$, the already fully filled $m$ flavour(s) 'unfill' partially and merge with the currently filling flavour, so that out of the four flavours there are $m + 1$ flavours with density $\nu_i = \nu/(m + 1)$ and $3 - m$ flavours with zero density $\nu_i = 0$. While there are certainly refinements that could be made on this model, this result shows a general feature of Coulomb-induced symmetry breaking in TBG: the charges of the full-filled flavours could still participate in further interactions when more charges are added.

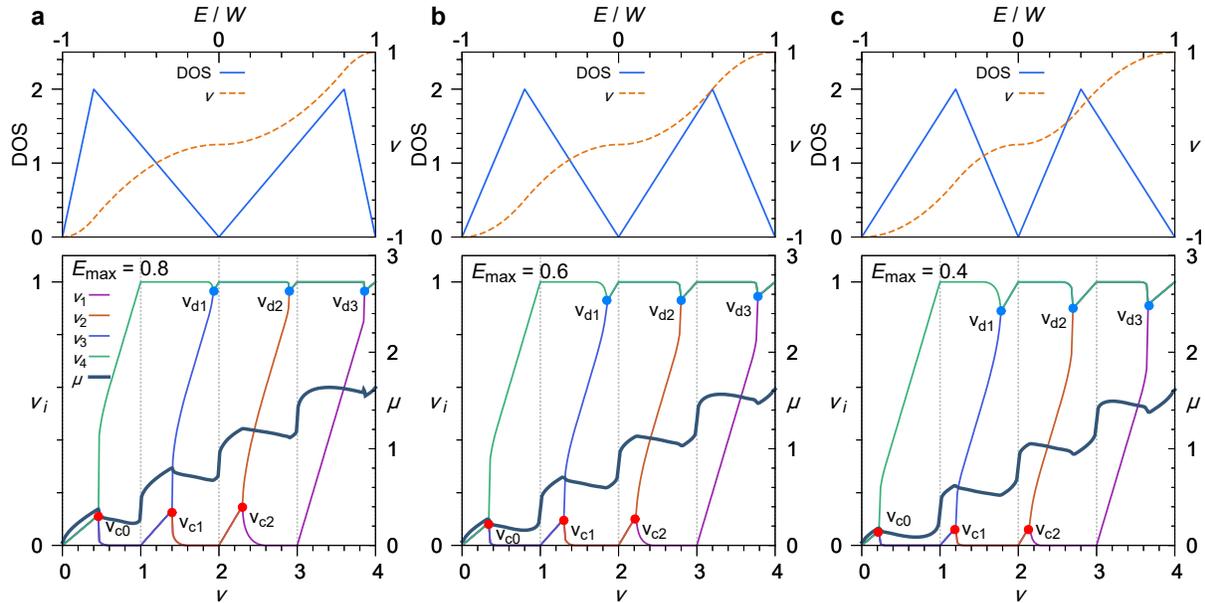

FIG. S6. Dependence of the flavour symmetry breaking on the DOS shape. Shown are the DOS and $\nu_i$ versus $\nu$ for three different values of $E_{\max}$, which is the position of the maximum in the DOS. The parameters are $U = 0.45$, $J = 0.35$.

At zero magnetic field, this model qualitatively explains our observed negative compressibility near all integer fillings, including before $\nu = 4$. However, it should be noted that so far this model does not have enough information to reproduce the in-plane magnetic field data, since we do not know the spin/valley texture of each flavour and how their energies evolve in a magnetic field. One could in principle attempt to include such effects by adding a Zeeman energy term $gB \sum_i \nu_i s_i$ into the free energy, where $s_i = \pm 1$ are the spins of the flavours (two up and two down) and $g$ is the $g$-factor. However in doing so, at $\nu = 2$ the system will preferably enter a ferromagnetic state to minimize the Zeeman energy term, which does not match our experimental observations. We will further discuss the implications of our experimental data at $\nu = 2$ in Section V.

### B. With Perpendicular Magnetic Field

The energy spectrum of TBG is quantized into Hofstadter subbands in a strong perpendicular magnetic field, as shown in Fig. 3c. The largest persisting gaps within the spectrum, besides the trivial gaps at the superlattice density, are the Chern number $C = \pm 1$ gaps shown in Fig. 3c, which are equivalent to the Landau level gap $\pm 1$ for a single flavour. The center band between the $C = 1$ and $C = -1$, which is evolved from the zeroth Landau level in graphene, has a Chern number of 2

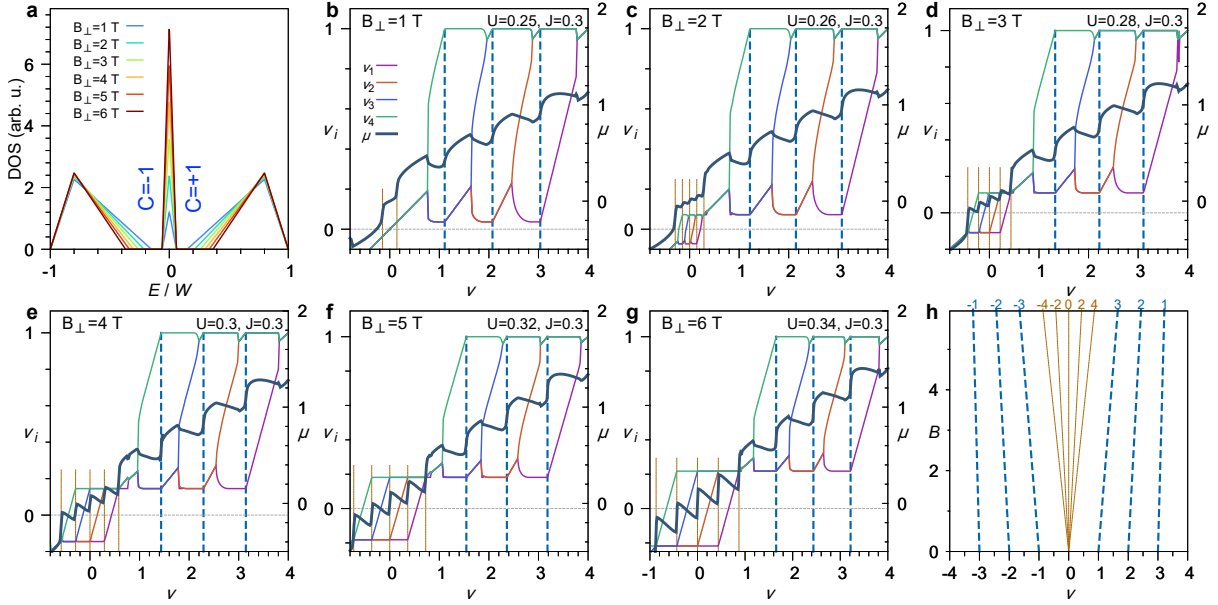

FIG. S7. flavour symmetry breaking and creation of correlated Chern gaps in a perpendicular magnetic field. (a) The single-particle DOS that captures the major features of a Hofstadter spectrum. The central peaks represent the equivalent of the zeroth Landau level, above and below which there are gaps with Chern numbers $\pm 1$ respectively. (b-g) flavour symmetry broken $\nu_i$ and $\mu$ at different perpendicular fields. $U$ is slightly increased with $B_\perp$ to ensure proper symmetry breaking that matches the experiment. The vertical brown and blue dashed lines denote the charge neutrality Landau level gaps and correlated Chern gaps respectively. (h) Positions of the Landau level gaps and correlated Chern gaps as a function of magnetic field, which clearly shows their Chern number as labeled on top of the plot, consistent with the Streda formula $C = \phi_0 \left( \frac{\partial n}{\partial B_\perp} \right)$ [16].

(from the two layers) and a density of $\Delta n = 2B_\perp/\phi_0$ (or $\Delta \nu = 2\phi/\phi_0$, where $\phi = 4B_\perp n_s^{-1}$ is the flux per unit cell).

We model the DOS in a similar manner as the zero field case while capturing the $\Delta C = 2$ Chern band at zero energy. Fig. S7a shows the model DOS we used for different magnetic fields. A triangular peak at $E = 0$ represents the DOS in the $C = 2$ Chern band, and the area under this peak is equal to $2\phi/\phi_0$. We set the gap between the peak and upper/lower triangular bands to be proportional to $\sqrt{B}$, as expected for Landau level gaps in graphene (behaviour of the actual gap size is more complicated, see Fig. 3c). While there are other gaps in the spectrum, such as the gaps before reaching the superlattice densities, they are often not observed in TBG near the magic-angle [9, 17–19], and it is therefore a reasonable approximation to neglect them in this simplified model.

With this model, we can simulate the behaviour of $\mu$ versus $\nu$ when $B_\perp$ is continuously varied.

We minimize the free energy $G$ in Eq. 12 within the bounds $-1 < \nu_i < 1$ using the DOS shown in Fig. S7a. Our results could account for both the symmetry-broken Landau levels and the correlated Chern insulator states in a unified way, as shown in Fig. S7b-f. Near the charge neutrality, above $B_\perp = 1\,\text{T}$ we observe symmetry breaking in the zeroth Landau level ($\Delta C = 2$ band), which gives rise to the quantum Hall ferromagnetism as in monolayer graphene [20, 21]. When the total filling $\nu$ is increased, the filling per flavour $v_i$ increases from -1 to 1 one flavour by one flavour. Note that when a magnetic field is present, zero density $\nu_i$ is not a 'stable' filling anymore. The stable fillings are now $\nu_i = \pm 1$ and $\nu_i = \pm\phi/\phi_0$, where the Chern numbers are $C = 0$ and $C = \pm 1$ respectively (see Fig. 3c and Fig. S7a). In the mean time, the chemical potential extrema that were near $\nu = m, m = 1, 2, 3$ now shift to higher $\nu$, again because the stable fillings are no longer $\nu = m$ but now $\nu = m + (4 - m)\phi/\phi_0$, corresponding to $m$ fully filled flavours and $4 - m$ flavours with chemical potential in the $C = +1$ gap. The Chern number of this gapped state is $(4 - m)C$, which is evident if we plot their positions versus $B_\perp$, since the Chern number is related to the slope by $C = \phi_0\left(\frac{\partial n}{\partial B_\perp}\right)$. These features are identified in our experiments, as shown in Fig. 3a. The chemical potential in Fig. S7b-g are also shown in Fig. 3b to compare with the experimental data side by side.

In our experimental data, we also found evidence for sub-meV gaps for Landau level filling factors $\nu_{LL} = \pm 1, \pm 3$, denoted by green bars in Fig. 3e. These further broken-symmetry features cannot be explained in the above model, because the Chern numbers of the gaps near the charge neutrality always differ by an even number when a flavour is filled from the $C = -1$ gap to $C = +1$ gap and therefore cannot produce odd-filling gaps. However it is not hard to account for these gaps by breaking further symmetries in the system. One possibility is a breaking of $C_3$ symmetry near the charge neutrality point, which was suggested in recent experiments [22, 23]. Theoretically, the broken $C_3$ symmetry allows an energy gap with $C = 0$ to be developed at the center of the zeroth Landau level[24]. The possible Chern numbers in the gaps are now $C = -1, 0, 1$ per flavour, and can give rise to odd filling factors when Coulomb repulsion is turned on.

### C. Calculation of the Hofstadter Spectrum and the Chern Number

To numerically calculate the change of the Chern number during the flavour symmetry breaking, we first use the continuum model for TBG in a magnetic field to calculate the Hofstadter spectrum shown in Fig. 3c. The detailed algorithm of the calculation follows [25] and [26]. We used $\theta = 1.8°$ for this calculation, but the qualitative features remain similar at the magic-angle. The spectrum

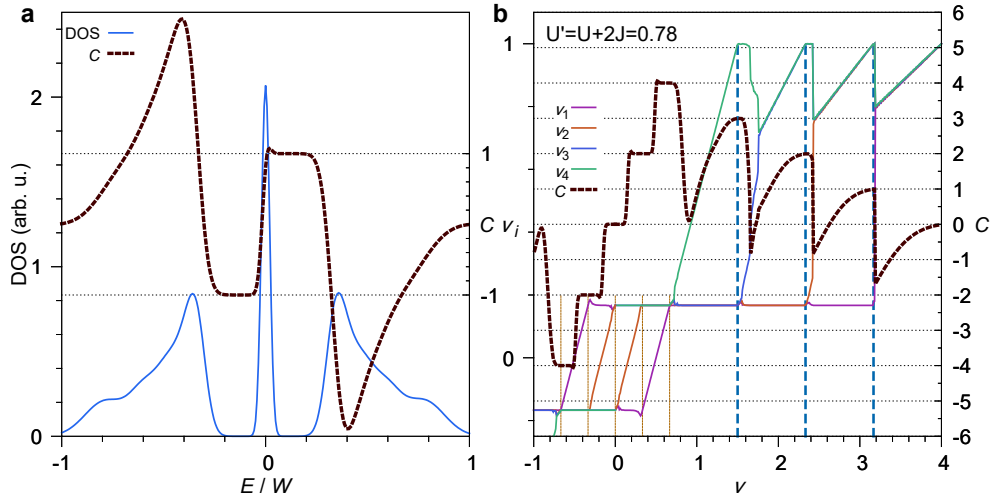

FIG. S8. Calculation of the Chern number in the correlated spectrum. (a) Single-particle DOS directly calculated from the Hofstadter spectrum of TBG, and the (per flavour) Chern number (total Hall conductivity in $e^2/h$). (b) flavour symmetry breaking calculation using the DOS in (a), and the total Chern number $C$ as the sum of Chern number from all four flavours.

is smoothed to account for possible twist angle disorder in the system, which yields a DOS where only the $C = \pm 1$ gaps above/below the zeroth Landau level are prominent, as is the case in our simplified model. The Chern number integrated below the Fermi level (*i.e.* Hall conductivity in terms of $e^2/h$) as a function of energy is directly calculated from the Hofstadter spectrum (see Ref. [24]) and smoothed by the same way as above. The smoothed single-particle DOS and Chern number are shown in Fig. S8a. One could also integrate the DOS and plot the same quantities versus the filling $-1 < \nu < 1$ (per flavour).

By minimizing the same form of free energy Eq. 12, we obtain the filling per flavour $\nu_i$ in the ground state and calculate the total Chern number as

$$C_{\text{tot}} = \sum_i C(\nu_i), \tag{21}$$

where $C(\nu_i)$ is the Chern number (Hall conductivity) per flavour as a function of the filling, as discussed in the previous paragraph. The result of this calculation is shown in Fig. S8b, as well as in Fig. 3d.

## V.  MAGNETIZATION AT THE $\nu = \pm 2$ STATES

In our measurement of the chemical potential in an in-plane magnetic field, we surprisingly find no dependence of $\mu$ on $B_\parallel$ near $\nu = \pm 2$, unlike near $\nu = \pm 1$, where $\mu$ shows a clearly spin-polarized

behaviour in a finite field. This observation is puzzling if one considers the evolution of the chemical potential as $\nu$ is gradually filled up from $\nu = 1$ to $\nu = 2$. If only one flavour is filled at a time (the other flavours are either fully filled or empty), the dependence of the chemical potential $\mu$ with $B_{\parallel}$ would be determined by the magnetic properties of that flavour, *i.e.* $\mu$ decreases (increases) as $g\mu_B B_{\parallel}$ if that flavour has spin aligned (anti-aligned) with $B_{\parallel}$. Since $\nu = 1$ has already a spin-polarized ground state with spin aligned with the magnetic field ($\mu$ decreases with $B_{\parallel}$), one may think that the second flavour being filled between $\nu = 1$ and $\nu = 2$ would have an anti-aligned spin if a spin-unpolarized state were to be obtained at $\nu = 2$, which was suggested by the transport experiments[9, 17, 27], as also shown in Extended Data Figure 2.

To attempt to resolve this dilemma, one possible scenario is that the flavour being filled is a mixture of spin-up and spin-down states, which does not have a net spin. Another possibility is that near $\nu = 2$, there are actually *two* flavours with opposite spins being filled at the same time. The latter case is in fact demonstrated in our theoretical model when the vHs in the single-particle DOS is closer to charge neutrality, as shown in Fig. S6c. Between $\nu = \nu_{d1}$ and $\nu = 2$, two flavours are simultaneously being filled while the other two flavours are empty. If these two flavours have opposite spin (or if they are both spin-unpolarized), the chemical potential would have no net response to the in-plane magnetic field.

However, our measurement of the magnetization, which is obtained by integrating the dependence of $\mu$ on $B_{\parallel}$ (see Methods), brings a bigger puzzle to the magnetic ground state of MATBG. While at $\nu = \pm 1$ (the $\nu = -1$ state is slightly shifted as discussed before) the magnetization does reach a value on the order $1\mu_B$ per moiré unit cell, at $\nu = \pm 2$ this magnetization does not reset to zero as one would expect for a spin-unpolarized state. While we wish to leave this as an open question, it is important to point out a new possibility that the $\nu = \pm 2$ state might have a more complicated ground state than previously thought, despite the apparent suppression of their transport gaps by the in-plane magnetic field.

## VI. DIFFUSIVITY AT OTHER FILLINGS

We further analyzed $\rho$, $\chi^{-1}$, and $D$ for other filling factors of interest. In particular we studied two regions. Firstly, near the CNP of MATBG, $\rho$ has a quadratic dependence in $T$ up to $T \sim 25\,\mathrm{K}$ (Fig. S9a), which is a signature of Fermi liquid behaviour. In this region, $\chi^{-1}$ exhibits a noticeable decrease with increasing $T$ for $\nu$ very close to the CNP, while the dependence diminishes as $\nu$ moves away from it, as shown in Fig. S9b. Fig. S9c shows the diffusivity (without removing residual

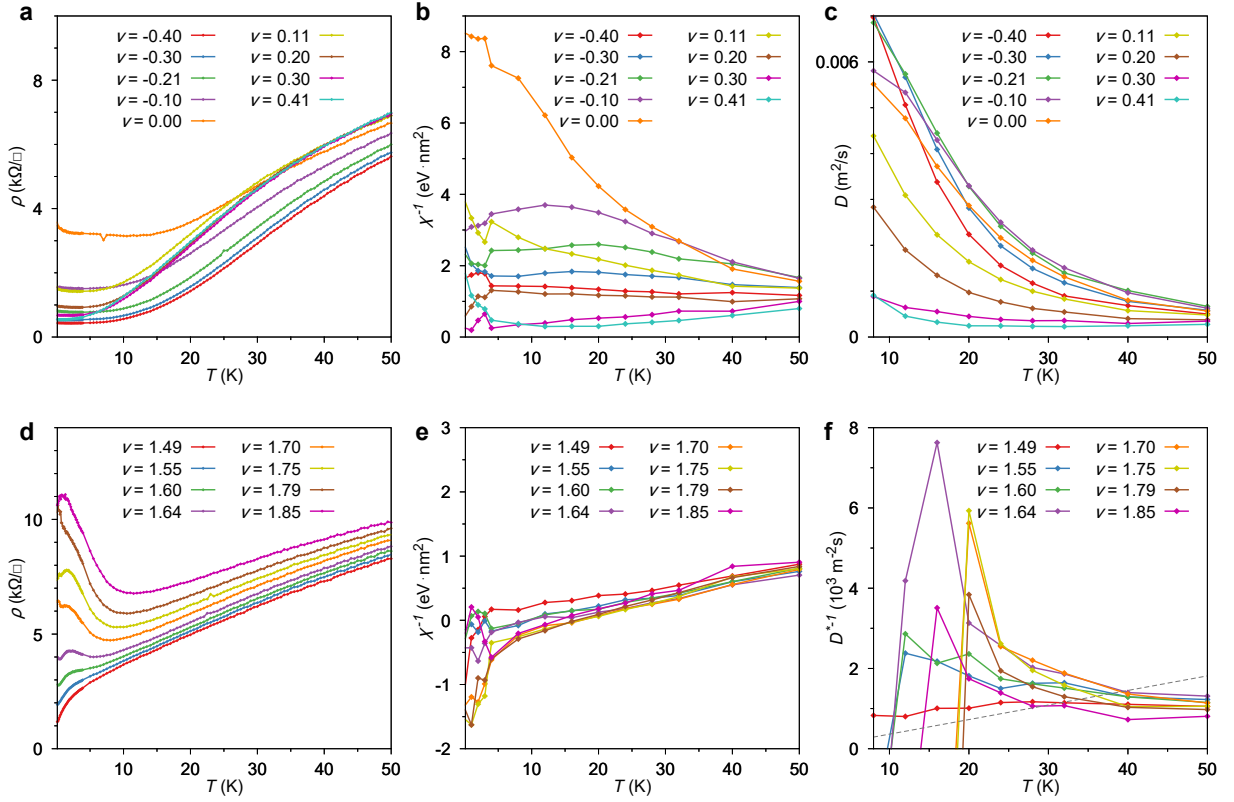

FIG. S9. Resistivity, inverse compressibility, and diffusivity at other filling factors. (a)-(c) Near MATBG CNP, $\rho \propto T^2$. $\chi^{-1}$ shows noticeable temperature dependence for $\nu$ nearer to CNP, and minimal dependence when it is away from it. (d)-(f) For $\nu$ across the region where $\chi^{-1}$ becomes negative near $\nu = 2$, linear $\rho - T$ is still observed. However, since $\chi^{-1}$ is linearly increasing from negative values to an asymptotic positive value at high temperatures, the resulting $D^{*-1}$ shows a diverging behaviour when $\chi^{-1}$ crosses zero.

resistivity) in this region.

Secondly, we look at the region of densities where the compressibility goes negative at low temperatures. Specifically, we study the filling factors near $\nu = 2$ where $\chi^{-1} < 0$ but $\rho$ still shows linear dependence in $T$, as shown in Fig. S9d. Since $\chi^{-1}$ crosses zero when going from negative values at low $T$ to positive values at high $T$ (Fig. S9e), the inverse effective diffusivity $D^{*-1}$ shows a diverging behavior near the zero-crossing temperature, as shown in Fig. S9f. $D^{*-1}$ violates the diffusivity bound $D_{\text{bound}} = \hbar v_F^2/(k_B T)$ during this divergence, but falls within the bound above around 30 K to 40 K. These observations raise an interesting question of how to theoretically interpret the Nernst-Einstein relation when $\chi^{-1}$ is negative, since it implies that the diffusivity is also negative. In fact, the presence of a negative diffusivity does not violate any fundamental thermodynamic law, as at large enough length scales the dynamics of density

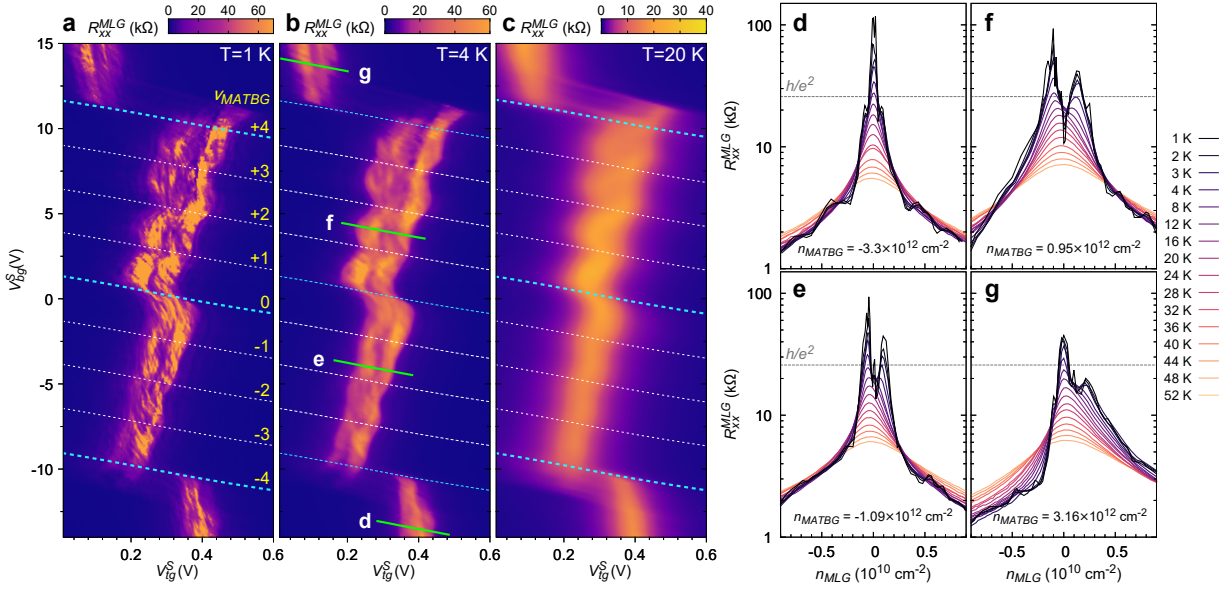

FIG. S10. Monolayer graphene resistance peak at charge neutrality point (CNP) splits in the presence of closeby MATBG. (a)-(c) Resistance of MLG CNP peak at temperatures $T = 1$ ,$4$ ,$20$ K, respectively. $x$-axis is $V_{tg}^S = V_{tg} + 0.03V_{bg} + 0.1$ and $y$-axis is $V_{bg}^S = V_{bg}$. Blue dash lines denote the superlattice density $\pm n_s$ and the charge neutrality point of MATBG, and white dash lines indicate the filling factors $\nu_{MATBG} = \pm 1, \pm 2, \pm 3$. Green lines show the constant $n_{MATBG}$ lines where scans (d)-(g) are taken. (d)-(g) Temperature dependence of the MLG CNP peak taken as a function of $n_{MLG}$ at constant $n_{MATBG}$.

fluctuation is always dominated by long-range Coulomb repulsions [28]. At short wavelengths, however, negative diffusivity associated with negative compressibility would lead towards phase separation in the charge density, possibly creating ordered phases like a Wigner crystal or a striped phase [29]. In such a regime, it is quite possible that the existing predictions of the diffusivity break down. We encourage further theoretical work to investigate the possible role of negative compressibility in the strange metal regime.

## VII. SPLITTING OF MLG DIRAC POINT IN THE MATBG FLAT BANDS

We noticed that at low temperatures, the resistance peak of MLG at charge neutrality shows a prominent splitting in $n_{MLG}$ when $n_{MATBG}$ is within the flat bands ($|n_{MATBG}| < n_s$), as shown in Fig. S10. When $n_{MATBG}$ is outside of $\pm n_s$, as shown in Fig. S10d and g, the resistance peak is one sharp peak at all temperatures. However, when $n_{MATBG}$ is within the flat bands, the resistance peak in MLG is clearly split into two peaks, as evident from the colour maps in Fig. S10a,b and linecuts in S10e,f. This behaviour disappears when temperature is raised to above about 20 K. At

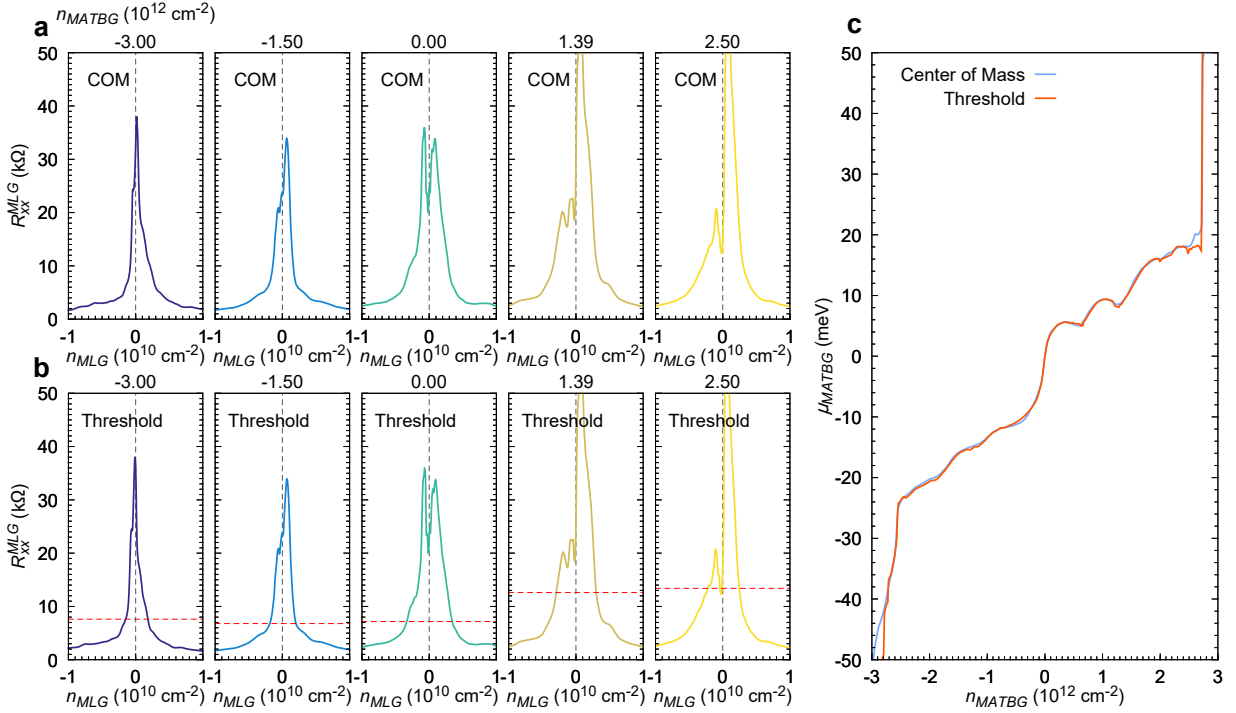

FIG. S11. Extraction of the peak positions. (a) $R_{xx}^{\mathrm{MLG}}$ peak centered at $n_{\mathrm{MLG}} = 0$ (vertical dashed line) using the center-of-mass (COM) method, at five representative densities of $n_{\mathrm{MATBG}}$. (b) Same data centered using the threshold method, with the threshold shown in red dashed lines. The threshold is chosen as 20% of the maximum resistance. (c) Comparison of $\mu_{\mathrm{MATBG}}$ extracted using the two methods.

all $n_{\mathrm{MATBG}}$, the MLG peak resistance keeps increasing as the temperature decreases down to 1K and reaches above $h/e^2$, suggesting that the Dirac point might be slightly gapped.

While we do not fully understand this behaviour at this point, our observation points towards an interaction-driven splitting of the resistance peak in MLG due to its close proximity to a flat-band metal. Clearly, the splitting cannot be explained purely by disorder in the MLG, as such disorder should result in a splitting regardless of the density in MATBG. Furthermore, the temperature dependence indicates that the disappearance of the splitting seem to be not just a thermal smearing of the two peaks, but instead a merger into a single peak around 10 K. This suggests that the splitting is a low-temperature phenomenon, and there might be an associated phase transition around 10 K. We hope that further experiments can clarify such puzzling features in our data.

To obtain the peak positions of MLG charge neutrality in presence of these splittings, which is essential for the extraction of $\mu_{\mathrm{MATBG}}$ (step 4 in Section II B), we can use two different methods, which give consistent results, as shown in Fig. S11. The first method uses the 'center of mass'

of the peak, *i.e.* a weighted average of $n_{\mathrm{MLG}}$ using the value of $R_{xx}^{\mathrm{MLG}}$ as the weights. In the second method, we first find the extent of the peak by looking for intersections of the resistance curve with a threshold, which is typically chosen as $20\,\%$ of the maximum resistance. We then take the average of $n_{\mathrm{MLG}}$ at the intersections to determine the center position. We find that the two methods give very similar results in general. The extracted $\mu_{\mathrm{MATBG}}$ using the two methods (Fig. S11c) are largely the same, with small discrepancies visible only near $\nu = +4$. For all our extractions in the figures in the main text, we used the first method for consistency.



\end{document}